\documentclass[aps,prb,reprint,superscriptaddress]{revtex4-2}

\usepackage[dvipdfmx]{graphicx,color}
\usepackage{amsmath}
\usepackage{amssymb}
\usepackage{enumerate}
\usepackage{textcomp}
\usepackage{ulem}

\newcommand{\mib}[1]{\mbox{\boldmath $#1$}}
\newcommand{\diffa}[1]
{{\rm d}{#1}}
\newcommand{\diffbaa}[3]
{\frac{{\rm d}^2{#1}}{{\rm d}{#2} {\rm d}{#3}}}
\newcommand{\dsp}{\displaystyle}

\begin{document}
	
	
\title{Realistic Simulation of Valence-band Photoemission Spectra
		by using First-principles Calculations on the Basis of FLAPW Method}
	
\author{Noriaki Hamada}\email[]{hamada@rs.tus.ac.jp}
\affiliation{Department of Physics, Tokyo University of Science, Noda, Chiba 278-8510, Japan}
\affiliation{Center for Spintronics Research Network, Osaka University, Toyonaka, Osaka 560-8531, Japan }
\author{Satoshi Nakamura}
\affiliation{Metro Inc., 1-11-1 Osaki, Shinagawa, Tokyo 141-0032, Japan}
\author{Kenta Takahashi}
\author{Mario Okawa}
\author{Tomohiko Saitoh}\email[]{t-saitoh@rs.tus.ac.jp}
\affiliation{Department of Applied Physics, Tokyo University of Science, Katsushika, Tokyo 125-8585, Japan}
	
	
\date{\today}
	
\begin{abstract}
		
In order to reproduce experimental photoemission spectroscopy (PES) data by using band structure calculations with linearized augmented-plane-wave method, we developed a new method to calculate the atomic-subshell photoelectron differential cross section (PE-DCS) of the valence band.
PE-DCS was calculated to be the sum of the products of the partial density of states (PDOS) and the photoionization differential cross section (PI-DCS) of each atomic subshell.
To construct an appropriate PDOS beyond the conventional muffin-tin sphere (MTS) PDOS, we proposed a ``nearly-neutral atomic sphere", which can be defined as a natural extension of MTS.
Although the calculational method for atomic-subshell PI-DCS was the standard Hartree-Fock-Slater method,
we developed a new computer code in order to deal with the unoccupied electron shells in the ground state of isolated atoms up to $\sim$10 keV.
Details of the PE-DCS calculations were examined by using two simple examples of Na metal and NaCl, and then, the method was applied to hard x-ray PES (HX-PES) data of CuAlO$_2$. The experimental spectra were satisfactorily reproduced, demonstrating the potential applications of our method to HX-PES spectra of many compounds.
\end{abstract}

\maketitle


\section{Introduction}

Photoemission spectroscopy (PES) is a powerful and the most direct experimental technique to investigate the electronic structure of solid-state materials.
In particular, valence-band PES spectra have fundamental importance in solid state physics because they provide us rich information on the electronic structure that is responsible for the physical properties in solids.

Because valence-band PES spectra of polycrystalline samples (meaning angle-integrated valence-band PES spectra \cite{Note2}) reflect the total electron density of states (DOS), 
they have long been compared with and analyzed by using first-principles band structure calculations, which are their best theoretical counterpart within the one-electron approximation.
In reality, however, the valence-band PES spectra as measured cannot directly be compared with theoretical total DOS
because the valence bands of (even elemental) solids consist of multiple atomic subshells of multiple elements, and the atomic-subshell photoionization cross sections differ from a subshell of an element to another \cite{yeh85,trzhaskovskaya01,trzhaskovskaya02}.
In theory, therefore, the valence-band PES process from a polycrystalline sample into a particular detection solid angle (with respect to the direction of the incident photons) should be described as the differential cross section of the total photoelectrons due to the incident photons from the valence band into this solid angle. In this paper, we call this quantity \textit{photoelectron differential cross section} (PE-DCS) \cite{Note3}.

In order to calculate PE-DCS, we can adopt a simple but quite effective method, which we call the \textit{atomic-limit model} in this paper. The atomic-limit model calculates PE-DCS as the sum of the product of the atomic-subshell \textit{photoionization differential cross section} (PI-DCS) and the partial DOS (PDOS) derived from a set of first-principles band structure calculations. 
Therefore, the calculated PE-DCS within this model is a direct manifestation of the atomic-orbital constituents in PES and this is one of the most important and useful features of the atomic-limit model.
In fact, because of its simple concept, photoemission experimentalists have traditionally constructed simulation PES spectra based on this model for many years by using a table of atomic-subshell PI-DCS's \cite{yeh85} and band structure calculations that they obtained from their theoretical collaborators. However, these analyses have been done empirically, without any concrete theoretical framework. For example, the O $2p$ PDOS in oxide compounds have been multiplied empirically by a factor of $\sim$3 \cite{Saitoh97}.

In this paper, we develop a concrete theoretical framework based on the linearized augmented-plane-wave (LAPW) method for calculating PE-DCS within the atomic-limit model and thus simulating valence-band PES spectra of polycrystalline samples and/or rough surfaces.
The reason we are targeting such samples is that so many materials with interesting properties are polycrystalline and/or hard to obtain their flat surfaces enough to perform angle-resolved PES (ARPES) measurements.

On using the atomic-limit model, two important conditions exist for the PDOS that have never been considered in the traditional analysis; 
(1) the sum of PDOS's must be equal to the total DOS, and
(2) the integration of PDOS at an atom up to the Fermi level must be equal to the valence-electron number of the neutral state of that atom as long as we deal with the atomic-subshell PI-DCS's of stand-alone neutral atoms.

In the APW method, it is typical that the PDOS is evaluated inside the muffin-tin sphere (MTS) and the sum of MTS-PDOS's is usually much smaller than the total DOS. This is because the MTS radius $r^{\rm mts}$ is as small as the core radius, and thus the most part of the \textit{s-p} wave functions is located outside the MTS. As a result, each PDOS is never ``neutral" nor is the sum of MTS-PDOS's equal to the total DOS, being incompatible with the two conditions.

Therefore, in order to realize the PDOS satisfying the two conditions as good an approximation as possible, we introduce a new concept ``nearly-neutral atomic sphere", which we call AS1 in this paper.
AS1 is characterized by the nuclear charge $Z$, the valence-electron number $N$, the valence-electron subshell $nl$, and the atomic sphere radius $r^{\rm as1}$.
For example, the AS1 of Na atom is defined by $Z=11$, $N=1$, $nl =$ $3s$ and $3p$, and 
$r_{\rm Na}^{\rm as1}\approx$ 4.3~\AA, 
the last of which will be determined later.
Here are two important points to be noted; 
(1) $r_{\rm Na}^{\rm as1}$ is much larger than 
$r_{\rm Na}^{\rm mts}\approx$ 2.3~\AA, and (2) the $3p$ subshell is included in the valence-electron subshell because it is fractionally occupied in crystal.

For the atomic-subshell PI-DCS, the Yeh-Lindau table \cite{yeh85} has been used for a long time over 35 years.
The table, however, provides the PI-DCS neither for the photon energy higher than 1500 eV, nor for the unoccupied valence-electron shell in the ground state of atom. 
Recent new tables by Trzhaskovskaya \textit{et al.} \cite{trzhaskovskaya01,trzhaskovskaya02,trzhaskovskaya18} provides the PI-DCS from 100 eV up to 10 keV, but still does not deal with the latter issue.
The recent WIEN2k deals with the underestimated PDOS
by using a similar concept of ``renormalization of PDOS'', but still employs the above tables \cite{Bagheri19,Blaha20}.
Therefore, we develop a new computer program to calculate the PI-DCS for any photon energy ($\lesssim 10$ keV)
and for any electron subshells \textit{including the unoccupied subshells} such as $3p$ of Na.

The organization of this paper is as follows:
In Section \ref{sec:method}, the method of calculations is presented in detail with simple examples of Na and NaCl.
In Section \ref{sec:CuAlO2r}, the method is applied to a copper delafossite oxide CuAlO$_2$ and the results are
compared with hard x-ray photoemission spectroscopy (HX-PES) measurements \cite{takahashi19}. Finally in section \ref{sec:conclusion}, concluding remarks are presented.

\section{Method\label{sec:method}}

In the photoemission process, the energy conservation law is written as
\begin{eqnarray}
  \hbar\omega = -\epsilon + W + K~,
\end{eqnarray}
where $\hbar\omega$ is the incident photon energy, $-\epsilon$ is the binding energy of the photoelectron, $W$ is the work function, and $K$ is the kinetic energy of the photoelectron. For the valence-electron photoemission in crystal, $\epsilon$ is identified as the quasiparticle energy, which can approximately be estimated by first-principles band structure calculations.
In a photoemission experiment, $\hbar\omega$ is fixed and $W$ can be treated as a constant value. Thus, PE-DCS at the incident photon energy $\hbar\omega$ ($P^{\hbar\omega}$) is written as
\begin{equation}
  P^{\hbar\omega}=\diffbaa{\sigma}{\Omega}{K}
  =\diffbaa{\sigma}{\Omega}{\epsilon}~,
\end{equation}
where 
$\sigma$ denotes the total photoelectron cross section and
$\Omega$ is the solid angle in the emitting direction of a photoelectron.

Based on the atomic-limit model, $P^{\hbar\omega}$ can be described as follows:
\begin{eqnarray}
P^{\hbar\omega}(\epsilon,\Theta)&=&
\sum_{\nu(k) n l}
P^{\hbar\omega}_{\nu n l}(\Theta)
D^{\rm AS}_{\nu(k) n l}(\epsilon)~.
\label{eq_PE-DCS}
\end{eqnarray}

\begin{figure}[htb!]
\centering
\includegraphics[width=5cm]{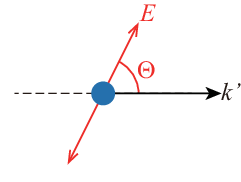}
\caption{\label{fig_Theta} (Color online) Geometrical relation between the polarization vector of incident photons $\mib{E}$ (red arrow) and the wave vector of a photoelectron $\mib{k'}$ (black arrow) from an atom (blue sphere). $\Theta$ is the angle between $\mib{E}$ and $\mib{k'}$.
The direction of the incident photon beam is perpendicular to the paper.
}
\end{figure}

In Eq.~(\ref{eq_PE-DCS}), $\dsp P^{\hbar\omega}_{\nu n l}(\Theta)$ is the atomic-subshell PI-DCS of the principal and the orbital angular momentum quantum numbers $n,l$ of an atomic element $\nu$.
$\Theta$ is the angle between the polarization vector of the incident photon and the wave vector of the photoelectron as shown in Fig.~\ref{fig_Theta}.
$\dsp D_{\nu(k) n l}^{\rm AS}(\epsilon)$ is the PDOS of the $nl$-subshell electrons at an atomic site labeled by $\nu(k)$, which is evaluated inside a certain atomic sphere.

Here, $\nu(k)$ denotes an atomic site in the primitive cell, distinguishing inequivalent sites $k$ of the atomic element $\nu$. 
Namely, $\nu(k)$ runs through all the inequivalent atomic sites (different elements and inequivalent sites of the same elements) in the primitive cell.
For example, because magnetite (Fe$_3$O$_4$) has two inequvalent Fe sites in the primitive cell, we denote their $3d$ PDOS as $\dsp D_{\text{Fe(1)}3d}^{\rm AS}(\epsilon)$ and $\dsp D_{\text{Fe(2)}3d}^{\rm AS}(\epsilon)$. If there are no inequivalent sites of the atomic element $\nu$ in the primitive cell, we omit $(k)$ for simplicity. Again, in the Fe$_3$O$_4$ case, the O $2p$ PDOS is described as $\dsp D_{\text{O}2p}^{\rm AS}(\epsilon)$.
Likewise, we always omit $(k)$ and use only $\nu$ for the quantities that depend only on atomic element such as the atomic-subshell PI-DCS $P^{\hbar\omega}_{\nu nl}(\Theta)$.

In this paper, we assume that the incident photons are linearly polarized, and all the calculations will be performed within the dipole approximation. In this situation, PI-DCS is only depend upon $\Theta$ regarding geometrical parameters.
The dipole approximation is quite appropriate for realistic comparisons with experiments because the quadrupole effects go
to zero in the typical experimental geometry of HX-PES measurement, in which strong $\Theta$ dependence appears due to high incident photon energy \cite{trzhaskovskaya01,Weinen15,takahashi19,Kobayashi09}.
Even if we use unpolarized photons, $\Theta$ is still very important and must be treated appropriately because the light is a transverse wave.

\subsection{AS1 radius and PDOS} 
\label{ssec:PDOS}

The general procedure to determine the AS1 radius of an atomic element $\nu$, $r_{\nu}^{\rm as1}$, as follows: (a) If $\nu$ forms a stable elemental crystal with metallic or covalent bonding, we calculate $r_{\nu}^{\rm as1}$ of that crystal structure. This is the case for all metal elements and Group 14 elements.
For example, $r_{\rm Cu}^{\rm as1}$ is determined by using fcc Cu metal.
(b) If $\nu$'s are other nonmetal elements, which usually become anions in solid-state compounds, we calculate the completely neutral atomic spheres of several typical compounds and determine a single value of $r_{\nu}^{\rm as1}$ as the average of them.
This is because these elements often form elemental molecular crystals in which the bonding is quite different from that in solid-state compounds.

PDOS, evaluated inside an atomic sphere in crystal,
is generally expected to properly reflect the electronic structure at the relevant atomic site. 
However, overlapping atomic spheres inevitably include the features of their nearest-neighbor sites. 
On the other hand, if we adopt a small atomic sphere radius to avoid overlapping, the magnitude of PDOS is significantly underestimated.
Thus, we have to realize these two conflicting requirements.
In the APW band structure calculations, the PDOS is usually evaluated inside MTS, which we denote as MTS-DOS, $D_{\nu(k) n l}^{\rm mts}(\epsilon)$.
Because $r^{\rm mts}$ is as small as the core radius, MTS-DOS is considered to properly reflect the local electronic states except for its magnitude. This means that the ``ideal" PDOS should have the same energy distribution as MTS-DOS and thus can be approximated by MTS-PDOS with some enhancement factor.

Therefore, we define
our new AS1-PDOS, $D_{\nu(k) n l}^{\rm as1}(\epsilon)$,
as a product of $D_{\nu(k) n l}^{\rm mts}(\epsilon)$ and the MTS-to-AS1 ehhancement factor $f_{\nu n l}$:
\begin{equation}
    D_{\nu(k) n l}^{\rm as1}(\epsilon)
    = f_{\nu n l}D_{\nu(k) n l}^{\rm mts}(\epsilon)~.
\label{eq_pdosas1}
\end{equation}
$f_{\nu n l}$ is defined as an atomic quantity with no energy dependence. Thus the energy distribution of $D_{\nu(k) n l}^{\rm as1}(\epsilon)$ is the same as that of $D_{\nu(k) n l}^{\rm mts}(\epsilon)$
except for the magnitude.
Meanwhile, it is very important that $f_{\nu n l}$ must include orbital dependence $n l$ because the radial extension of the local wave functions depend on $n l$.
To evaluate $f_{\nu n l}$, we adopt the radial wave function of an isolated atom $\nu$, $\varphi_{\nu n l}(r)$.
The partial norm of $\varphi_{\nu n l}(r)$ is defined as
\begin{equation}
  n_{\nu n l}(r)
  =\int_{0}^{r}r'^2\diffa{r'}\left| \varphi_{\nu n l}(r')\right|^2~,
  \label{eq_pnorm}
\end{equation}
and therefore $f_{\nu n l}$ is defined as
\begin{eqnarray*}
  f_{\nu n l}=\frac{n_{\nu n l}(r_{\nu}^{\rm as1})}
                   {n_{\nu n l}(r_{\nu}^{\rm mts})}~,
  \label{eq_enhancef}
\end{eqnarray*}
where $r_{\nu}^{\rm as1}$ and $r_{\nu}^{\rm mts}$ are 
the AS1 radius and the MTS radius at an atomic site $\nu$, respectively. $r_{\nu}^{\rm as1}$ is an important parameter that will be determined below.
It should be emphasized that $f_{\nu n l}$ depends on the electronic structure only of the isolated atom but not of the atom in a crystal.
Owing to this definition, the method has potential applications to many compounds with different crystal structures.

We expect
\begin{equation}
    D(\epsilon) \approx \sum_{\nu(k) n l}^{\rm [as1]} 
      D_{\nu(k) n l}^{\rm as1}(\epsilon)~,
    \label{eq_pdos_sum}
\end{equation}
where $D(\epsilon)$ is the total DOS.
In Eq.~(\ref{eq_pdos_sum}), we impose an important restriction; 
the summation of $nl$ is restricted to the valence-electron subshell. For example, only $3s$ and $3p$ orbitals in the case of Na metal, or only $3d$, $4s$, and $4p$ orbitals in the case of Cu metal are considered in Eq.~(\ref{eq_pdos_sum}).
The AS1-PDOS concept always includes this restriction for $nl$,
which is a sharp difference from the MTS-PDOS in which $l=0, 1, 2,3, \cdots$. We call the restriction ``AS1 summation" and it is explicitly denoted as ``[as1]" on the summation symbol.
  
We define the orbital occupation number of the atomic subshell $nl$ at one atomic site of $\nu(k)$ as
\begin{equation}
N_{\nu(k) n l}^{\rm as1}
=
\frac{1}{g_{\nu(k)}}
\int_{-\infty}^{\epsilon_{\rm F}}\diffa{\epsilon}\, 
D_{\nu(k) n l}^{\rm as1}(\epsilon)~,
\end{equation}
where $g_{\nu(k)}$ denotes the number of atoms labeled by $\nu(k)$ and $\epsilon_{\rm F}$ is the Fermi energy.
We also expect 
\begin{equation}
N_{\nu} \approx N_{\nu(k)}^{\rm as1}
\equiv
\sum_{n l}^{\rm [as1]} N_{\nu(k) n l}^{\rm as1}~,
\label{eq_sum1}
\end{equation}
where $N_{\nu}$ is the valence-electron number of one atomic element $\nu$ and $N_{\nu(k)}^{\rm as1}$ is the valence-electron number inside AS1 at one atomic site of $\nu(k)$ \cite{Note6}.

In the full-potentilal APW method, $r_{\nu}^{\rm mts}$ is just a technial parameter, because the accuracy of calculations
can always be kept good if the number of plane waves is taken appropriately.
Therefore,  $r_{\nu}^{\rm mts}$ can be fixed and
only $r_{\nu}^{\rm as1}$
is the unknown parameter to be determined.
We try to determine a single value of $r_{\nu}^{\rm as1}$ for each atomic element $\nu$ because a common ``$r_{\nu}^{\rm as1}$ table" independent of materials can make the AS1 concept a new strong method for understanding the electronic structure of a wide variety of materials.

If we impose the strict condition
\begin{equation}
    N_{\nu} = N_{\nu(k)}^{\rm as1}
    \label{eq_sum2}
\end{equation}
for Eq.~(\ref{eq_sum1}), we can determine $r_{\nu}^{\rm as1}$.
This is, however, not always appropriate because there exists a case that the most or even all electrons in crystal may not belong to the constituent atoms.
For example, in a group of materials called electride, some electrons by themselves behave as if they were an anion without any nucleus at their center \cite{inoshita14,inoshita15}.
Therefore, we determine the values of $r_{\nu}^{\rm as1}$ 
by using Eq.~(\ref{eq_sum1}) but as close to Eq.~(\ref{eq_sum2}) as possible.

Now, we are ready to present two test calculations about Na atom, namely, Na metal and NaCl. Na atom is suitable for examining the AS1 method because much of the $3s$ and $3p$ electron density exists outside MTS.
The AS1 summation of Na metal is taken for Na $3s$ and $3p$.
We adopt $r_{\rm Na}^{\rm as1}=4.3a_0$
deduced from bcc Na. Then,
we obtain $f_{\text{Na}3s}=5.5$ and $f_{\text{Na}3p}=9.9$ for $r_{\rm Na}^{\rm mts}=2.3a_0$, where $a_0$ is the Bohr radius ($\approx$ 0.529 {\AA}).

\begin{figure}[t]
\centering
\includegraphics[width=8.6cm]{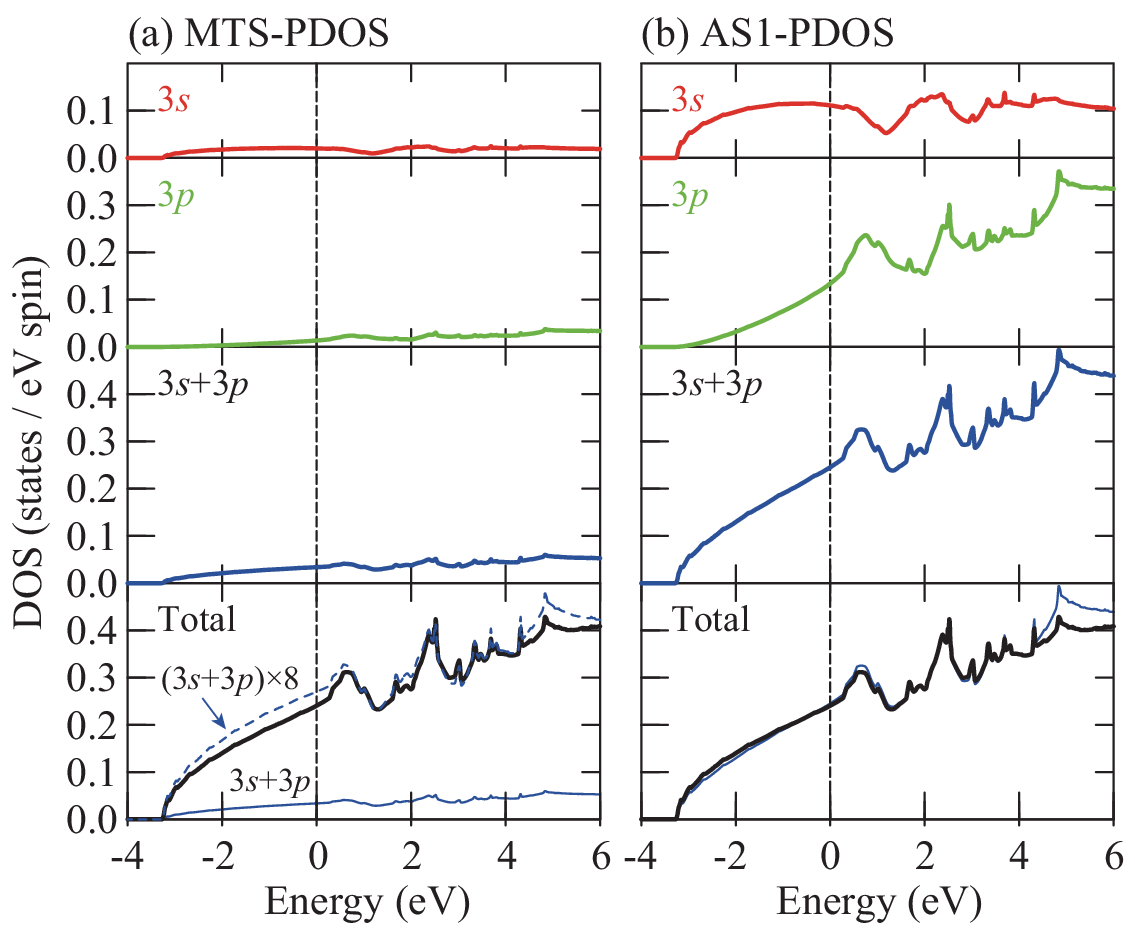}
\caption{\label{fig_PDOS_Na} (Color online)
PDOS of Na $3s$ and $3p$, the sum of those, and the total DOS of Na metal calculated by using (a) MTS and (b) AS1. The zero energy corresponds to $\epsilon_{\rm F}$.
Note that the both total DOS's are identical.
}
\end{figure}

Figure \ref{fig_PDOS_Na} shows a comparison of PDOS calculations on Na metal by using (a) MTS and (b) AS1, together with the total DOS. Note that the total DOS's in Panel (a) and (b) are identical.
It is obvious that
$D_{\text{Na}3s}^{\rm as1}(\epsilon) + D_{\text{Na}3p}^{\rm as1}(\epsilon)$ 
almost completely reproduces the total DOS except for the high energy region above 4 eV (the bottom figure of Panel (b)), while 
$D_{\text{Na}3s}^{\rm mts}(\epsilon) + D_{\text{Na}3p}^{\rm mts}(\epsilon)$ severely underestimates the total DOS by a factor of $\sim$1/8 (the bottom figure of Panel (a)).
Even if the MTS-PDOS is multiplied by a factor of $\sim$8, it cannot reproduce the total DOS as good as the AS1-PDOS, as shown in the bottom figure of Panel (a). This is because of the values of $f_{\text{Na}3s}$ and $f_{\text{Na}3p}$ are different, which is clearly demonstrating the importance of the $nl$ dependence in $f_{\nu n l}$.

\begin{figure}[t]
\centering
\includegraphics[width=8.6cm]{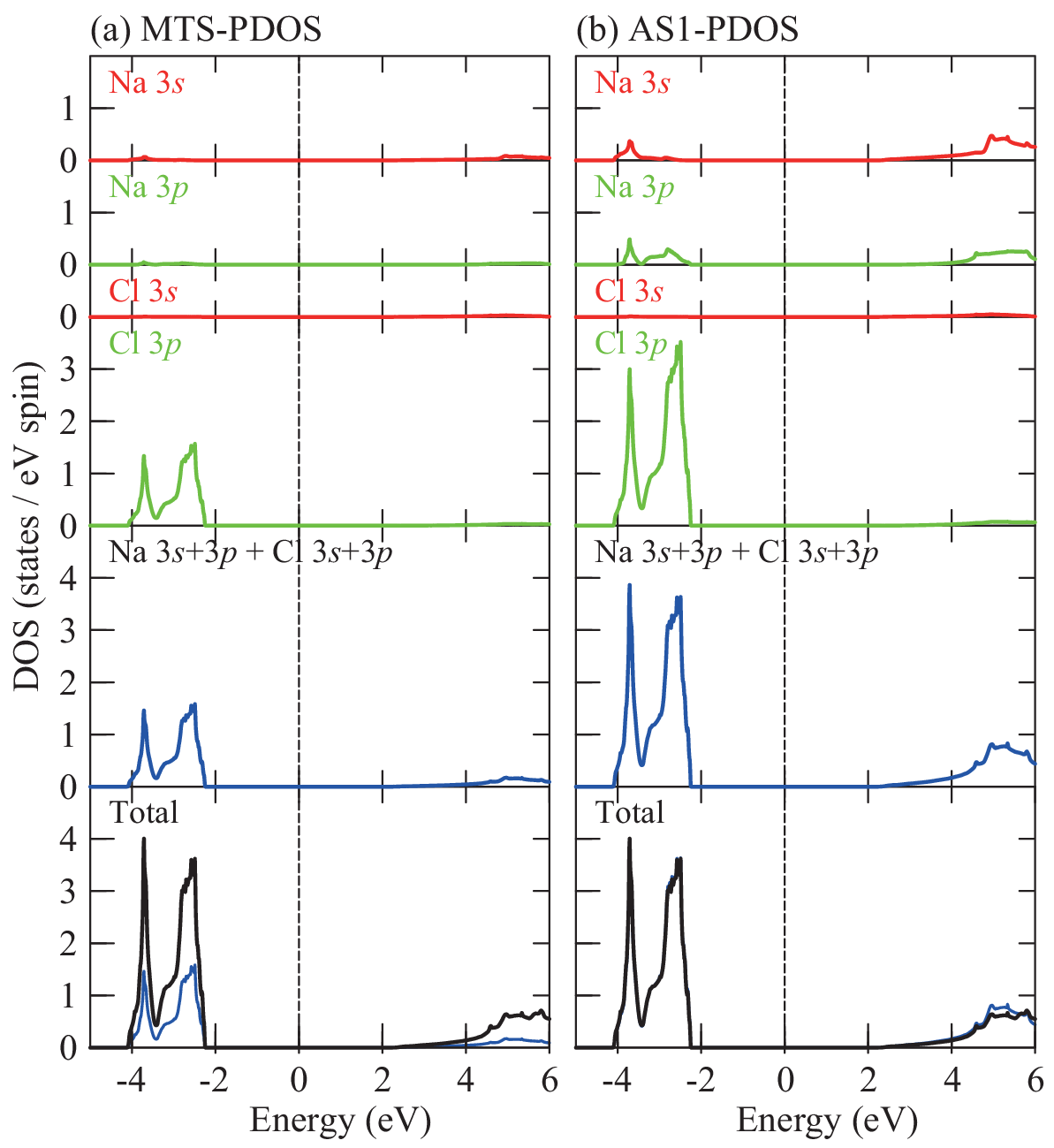}
\caption{\label{fig_PDOS_NaCl} (Color online)
PDOS of Na $3s$, $3p$, Cl $3s$, and Cl $3p$, the sum of those, and the total DOS of NaCl ionic crystal calculated by using (a) MTS and (b) AS1.
The zero energy corresponds to $\epsilon_{\rm F}$. Note that the both total DOS's are identical.
}
\end{figure}

Figure \ref{fig_PDOS_NaCl} shows a comparison of PDOS calculations on NaCl by using (a) MTS and (b) AS1, together with the total DOS. Again, the total DOS's in Panel (a) and (b) are identical. 
The AS1 summation of NaCl is taken for Na $3s, 3p$ and Cl $3s, 3p$. We adopt $r_{\rm Cl}^{\rm as1}=4.2a_0$
deduced by using the general procedure and
we obtain $f_{\text{Cl}3s}=1.6$ and $f_{\text{Cl}3p}=2.2$ for $r_{\rm Cl}^{\rm mts}=1.6a_0$.
In the bottom figure of Panel (b), the total DOS is reproduced so well by the sum of AS1-PDOS of Na $3s, 3p$ and Cl $3s, 3p$ that it is very hard to distinguish between the two except for the high energy region above 4 eV.
On the other hand, the sum of the four MTS-PDOS's underestimates the total DOS by a factor of $\sim$1/2.5 and the relative peak height of the two major peaks at $-2.5$ and $-3.7$ eV deviates from that of the total DOS (the bottom figure of Panel (a)).

\begin{table}[htb!]
	\caption{\label{tbl_f_Na}
		MTS radius $r_{\nu}^{\rm mts}$, 
		AS1 radius $r_{\nu}^{\rm as1}$, 
		MTS-to-AS1 enhancement factor $f_{\nu nl}$,
		MTS orbital occupation number $N_{\nu nl}^{\rm mts}$, and
		AS1 orbital occupation number $N_{\nu nl}^{\rm as1}$
		in Na metal.
		The number in the parenthesis is  
		$N_{{\rm Na}3s}^{\rm mts} + N_{{\rm Na}3p}^{\rm mts}$.
	}
	\begin{ruledtabular}
		\begin{tabular}{|c|cccc|}
			$\nu$ & \multicolumn{4}{c|}{Na} \\ 
			$r_{\nu}^{\rm mts}/a_0$ & \multicolumn{4}{c|}{2.3} \\ 
			$r_{\nu}^{\rm as1}/a_0$ & \multicolumn{4}{c|}{4.3} \\ 
			\hline
			$nl$ & $3s$ & $3p$ & 
			$N_{\rm Na}^{\rm as1}$ & $N_{\rm Na}$ \\ 
			\hline
			$f_{\nu nl}$ & 5.5 & 9.9 &  &  \\ 
			$N_{\nu nl}^{\rm mts}$ & 0.11 & 0.03 & (0.14) &  \\  
			$N_{\nu nl}^{\rm as1}$ & 0.62 & 0.34 & 0.96 & 1   
		\end{tabular}
	\end{ruledtabular}
\end{table}

Tables \ref{tbl_f_Na} and \ref{tbl_f_NaCl} summarize the used parameters and the calculated orbital-occupation numbers.
Reflecting Fig.~\ref{fig_PDOS_Na},
$N_{{\rm Na}3s}^{\rm mts}$ and $N_{{\rm Na}3p}^{\rm mts}$ in particular are very small as shown in Tables \ref{tbl_f_Na}.
$N_{{\rm Na}3s}^{\rm mts}$ is only 0.11 electrons although the whole one electron of $N_{\rm Na}$ originates from the Na $3s$ orbital in an isolated Na atom.
As a result, there are only 0.14 electrons inside $r_{\rm Na}^{\rm mts}$.
The number of Na $3s$ electrons recovers up to 0.62 in $N_{{\rm Na}3s}^{\rm as1}$ but still fewer than $N_{\rm Na}$, resulting in the fact that the Na $3p$ orbital is occupied by 0.34 electrons. $N_{\rm Na}^{\rm as1}$ is, thus, 0.96 close to $N_{\rm Na}$.

For NaCl ionic crystal in Table \ref{tbl_f_NaCl}, the Na $3p$ orbital is occupied by 0.53 electrons, which is much larger than the Na $3s$ occupation of 0.28. Because the Na $3p$ orbital is originally unoccupied in an isolated atom,
these Na $3p$ electrons are actually located at the neighbor sites and in the interstitial region, reflecting the ionic character of Na in NaCl. 
Therefore, the orbital-occupation numbers can be a rough measure of the bond character, particularly for cations.
We note that such analyses on AS1 calculations can be realized by determining the common value of $r_{\nu}^{\rm as1}$ for different materials.

\begin{table}
\caption{\label{tbl_f_NaCl}
MTS radius, $r_{\nu}^{\rm mts}$, 
AS1 radius, $r_{\nu}^{\rm as1}$,  
MTS-to-AS1 enhancement factor, $f_{\nu nl}$, 
MTS orbital occupation number, $N_{\nu nl}^{\rm mts}$, and
AS1 orbital occupation number, $N_{\nu nl}^{\rm as1}$ 
for $\nu=$ Na and Cl in NaCl crystal.
The numbers in the parentheses are  
$N_{{\rm Na}3s}^{\rm mts} + N_{{\rm Na}3p}^{\rm mts}$
and
$N_{{\rm Cl}3s}^{\rm mts} + N_{{\rm Cl}3p}^{\rm mts}$.
}
\begin{ruledtabular}
\begin{tabular}{|c|cccc|cccc|}
$\nu$ &
\multicolumn{4}{c|}{Na} & \multicolumn{4}{c|}{Cl} \\
$r_{\nu}^{\rm mts}/a_0$ & 
\multicolumn{4}{c|}{2.3} & \multicolumn{4}{c|}{1.6} \\
$r_{\nu}^{\rm as1}/a_0$ &
\multicolumn{4}{c|}{4.3} & \multicolumn{4}{c|}{4.2} \\ 
\hline
$nl$ & 
$3s$ & $3p$ & $N_{\rm Na}^{\rm as1}$ & $N_{\rm Na}$ & 
$3s$ & $3p$ & $N_{\rm Cl}^{\rm as1}$ & $N_{\rm Cl}$ \\ 
\hline
$f_{\nu nl}$ &
5.5 & 9.9 &   &  & 1.6 & 2.2 &  &  \\ 
$N_{\nu nl}^{\rm mts}$&
0.05 & 0.05 & (0.10) &  & 1.16 & 2.33 & (3.49) &  \\
$N_{\nu nl}^{\rm as1}$&
0.28 & 0.53 & 0.81& 1 & 1.89 & 5.22 & 7.12 & 7
\end{tabular}
\end{ruledtabular}
\end{table}

\subsection{Atomic-subshell PI-DCS}

In the dipole approximation, the atomic-subshell PI-DCS is written as \cite{yeh85}
\begin{eqnarray}\label{eqn_pi-dcs}
P^{\hbar\omega}_{\nu n l}(\Theta)
&=& 
\frac{\sigma_{\nu n l}^{\hbar\omega}}{4\pi}
\left[1+\beta_{\nu n l}^{\hbar\omega}P_2(\cos \Theta)\right]~,
\end{eqnarray}
where $\sigma_{\nu n l}^{\hbar\omega}$ is
the (total) atomic-subshell photoionization cross section (PI-CS) per one electron for each atomic subshell, 
$\beta_{\nu n l}^{\hbar\omega}$ is the dipole coefficient, and $P_2(x)$ is the second-order Legendre polynomial,
$P_2(\cos \Theta)=(3\cos^2\Theta-1)/2$.
We note that our definition of $\sigma_{\nu nl}^{\hbar\omega}$ is per one electron while the standard definition of $\sigma_{n l}^{\hbar\omega}$ in Ref \cite{yeh85} is for each atomic subshell including the subshell electron number. This is because the subshell electron number is already included in PDOS's, see the definition of PE-DCS in Eq.~(\ref{eq_PE-DCS}).

The atomic-subshell PI-CS is evaluated by
\begin{widetext}
\begin{equation}
\sigma_{\nu n l}^{\hbar\omega}= \frac{4\pi^2\alpha}{3(2l+1)}
\frac{\hbar\omega}{\hbar\omega+\epsilon_{\nu n l}}
\left[ l\left(R^{k',l-1}_{\nu n l}\right)^2
     +(l+1)\left(R^{k',l+1}_{\nu n l}\right)^2 \right]~,
\end{equation}
and the dipole coefficient is
\begin{equation}
\beta_{\nu n l}^{\hbar\omega}
=
\frac{l(l-1)\left(R^{k',l-1}_{\nu n l}\right)^2
     +(l+1)(l+2)\left(R^{k',l+1}_{\nu n l}\right)^2
     -6l(l+1)R^{k',l-1}_{\nu n l}R^{k',l+1}_{\nu n l}
      \cos(\delta^{k',l-1}_{\nu n l}-\delta^{\nu,k',l+1}_{\nu n l})}
      {(2l+1)\left[ l\left(R^{k',l-1}_{\nu n l}\right)^2
     +(l+1)\left(R^{k',l+1}_{\nu n l}\right)^2 \right]}~,
\end{equation}
\end{widetext} 
with
\begin{equation}
R^{k',l'}_{\nu n l}=\int_0^{\infty}r^2\diffa{r}
\tilde{\varphi}_{\nu k'l'}(r)\,r\,\varphi_{\nu n l}(r)~.
\label{eq_intR}
\end{equation}

Here, $\varphi_{\nu n l}(r)$ is the $nl$ radial wave function of $\nu$ atom,
$\tilde{\varphi}_{\nu k'l'}(r)$ is the radial wave function of the photoelectron with $k'$ being the wave vector in $r\to\infty$, and $l'$ being the angular-momentum quantum number. $\delta^{\nu,k',l'}_{\nu n l}$ is the phase shift in $r\to\infty$. The wave vector, $k'$ is defined by
\begin{equation}
  \frac{\hbar^2k'^2}{2m}=\hbar\omega + \epsilon_{\nu n l},
\end{equation} 
with $\epsilon_{\nu n l}$ ($<0$) being the energy level 
of the initial electron state measured from the vacuum level. 
The ground state of the atom is calculated self-consistently with an appropriate electron configuration by using the Slater exchange potential. The total potential is assumed to be spherical and is continuously connected to 
$e^2/(4\pi\varepsilon_0 r)$
in $r\ge r_{\rm coul}$ with  $r_{\rm coul}$ being typically 5 {\AA}.
The same potential is used for calculating the photoelectron state. For the detail of calculational scheme, please refer to papers of Cooper and Zare \cite{cooper68}, Yeh and Lindau \cite{yeh85}, 
Cooper \cite{cooper93} and therein.

We developed a computer code to calculate
$\sigma_{\nu n l}^{\hbar\omega}$ and 
$\beta_{\nu n l}^{\hbar\omega}$ in the non-relativisitc approximation.
The calculational scheme is exactly the same as that of Yeh and Lindau \cite{yeh85}. 
Nevertheless, developing a new program code to calculate 
PI-CS and $\beta_{\nu nl}^{\hbar\omega}$ is still needed
because the PI-CS and $\beta_{\nu nl}^{\hbar\omega}$ of
the atomic orbitals that are unoccupied in a free atom but partially occupied inside crystals have not been included in the preceeding studies \cite{scofield73,yeh85,trzhaskovskaya01,trzhaskovskaya02,trzhaskovskaya18}.
 
In Table \ref{tbl_phicsbetaAc_Na}, our numerical results on the Na atom are compared with those of Yeh and Lindau for three photon energies of 80 eV, 320 eV, and 1500 eV.
The agreement of the two calculations is very good, generally more than 3 digits for $\hbar\omega\le$ 1500 eV.
However, the data at $\hbar\omega =$ 8047.8 eV in the Yeh and Lindau table may have very large numerical errors because their data are sometimes an order of magnitude different from our results.

In the hard x-ray region, the PI-CS calculation becomes harder with increasing photon energy, because the photoelectron wave length becomes much smaller than the valence-electron wave length.
In our calculations, the integration of Eq. (\ref{eq_intR}) is carefully performed by using the automatic mesh method for keeping the accuracy well.

\begin{table}[htb!]
\caption{\label{tbl_phicsbetaAc_Na}
Atomic-subshell PI-CS per electron $\sigma_{\nu n l}^{\hbar\omega}$ and the dipole coefficient $\beta_{\nu n l}^{\hbar\omega}$ for each atomic orbital of Na atom
calculated for three photon energies 80, 320, and 1500 eV.
Our results (present) are compared with those of Yeh and Lindau \cite{yeh85} (YL).
For the $s$ orbitals, $\beta_{\nu n \rm s}^{\hbar\omega}$ is strictly 2 in the non-relativistic approximation.}
\begin{ruledtabular}
\begin{tabular}{|c|cccc|cc|}
& 
\multicolumn{4}{c|}{$\sigma^{\hbar\omega}_{\nu n l}$ [kb]} &
\multicolumn{2}{c|}{$\beta^{\hbar\omega}_{\nu n l}$} \\
$\nu$ &
\multicolumn{4}{c|}{Na} & \multicolumn{2}{c|}{Na} \\ 
$nl$ & {$2s$} & $2p$ & {$3s$} & $3p$ & $2p$ & $3p$ \\
\hline
$\hbar\omega=$ 80 eV &  &  &  &  &  &  \\
present & {301.} & 1133. & {19.8} & 12.9 & 1.243 & 1.507 \\
YL & {301.} & 1134. & {19.8} & - & 1.243 & - \\
\hline
$\hbar\omega=$ 320 eV&  &  &  &  &  &  \\
present & {72.3} & 52.2 & {2.36} & 0.603 & 1.419 & 1.422 \\
YL & {72.1} & 52.3 & {2.36} & - & 1.420 &  - \\
\hline
$\hbar\omega=$ 1500 eV &  &  &  &  &  &  \\
present & {2.81} & 0.424 & {0.0853} & 0.00525 & 0.783 & 0.768 \\
YL  & {2.84} & 0.421 & {0.0841} & - & 0.786 & -    
\end{tabular}
\end{ruledtabular}
\end{table}

\subsection{PE-DCS and Photoemission spectrum}

In our atomic-limit model with the AS1 method, the PE-DCS $P^{\hbar\omega}(\epsilon,\Theta)$ is finally given by
\begin{eqnarray}
P^{\hbar\omega}(\epsilon,\Theta)
&=&
\sum_{\nu(k) n l}^{\rm [as1]}
P^{\hbar\omega}_{\nu n l}(\Theta)
D_{\nu(k) n l}^{\rm as1}(\epsilon)~.
\label{eqn_PE-DCS}
\end{eqnarray}

Using $P^{\hbar\omega}(\epsilon,\Theta)$ and the detection solid angle of photoelectrons $\Delta\Omega$,
the PES spectrum normalized by the incident photon flux density
$I^{\hbar\omega}(\epsilon,\Delta\Omega)$
is calculated as
\begin{eqnarray} %
I^{\hbar\omega}(\epsilon,\Delta\Omega)
&=&
\int_{\Delta\Omega}\diffa{\Omega}\,P^{\hbar\omega}(\epsilon,\Theta) \\
&\approx&
P^{\hbar\omega}(\epsilon,\bar{\Theta}) \Delta\Omega
\label{eqn_PES}
\end{eqnarray}
with the average of $\Theta$,
\begin{equation}
\bar{\Theta}
=
\frac{1}{\Delta\Omega}
\int_{\Delta\Omega}\diffa{\Omega}\,\Theta ~.
\end{equation}

If $\Delta\Omega$ is small enough with respect to the $\Theta$ angle, Eq.~(\ref{eqn_PES}) is a good approximation and
the PE-DCS can directly be compared with the experimental PES spectra.
The small $\Theta$-variation condition is actually realized in the typical experimental geometry of HX-PES \cite{trzhaskovskaya01,Weinen15,takahashi19,Kobayashi09}.
Therefore, we hereafter denote $\bar{\Theta}$ as $\Theta$ for simplicity.

\begin{figure}[htb!]
\centering
\includegraphics[width=8.6cm]{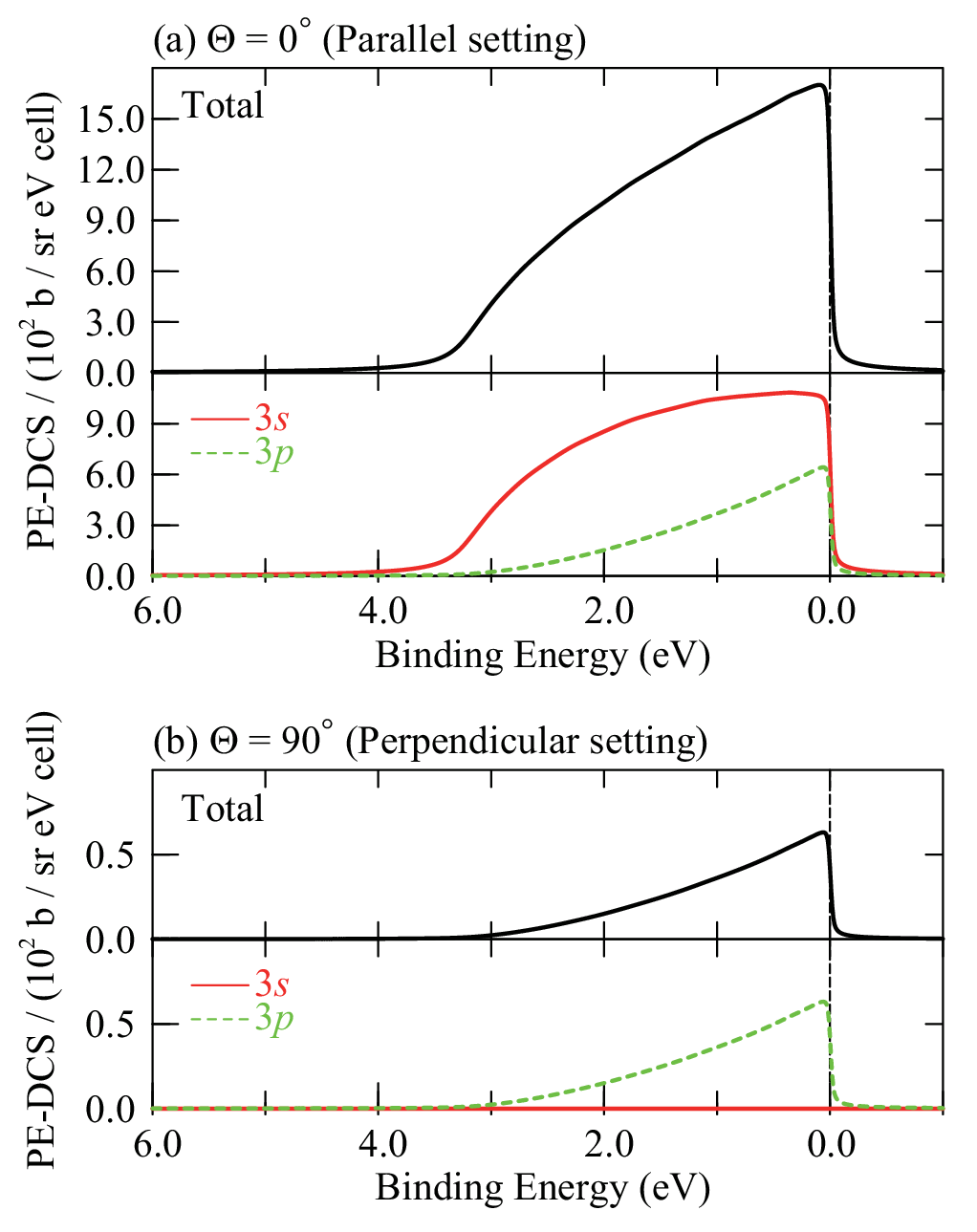}
\caption{\label{fig_pes80eV_Na} (Color online)
PE-DCS of Na metal at $\hbar\omega=80$ eV for (a) $\Theta = 0${\textdegree} and (b) $\Theta = 90${\textdegree}.
The zero binding energy corresponds to the Fermi energy.
The experimental resolution is set to be 30 meV. The Lorentzian broadening due to life time effect is also considered 
\cite{Saitoh97,takahashi19,Iwasawa09,Note4}.
}
\end{figure}

The $\Theta$ dependence of PE-DCS is a powerful tool to elucidate the atomic-orbital origins in PES spectra.
Figure~\ref{fig_pes80eV_Na} shows PE-DCS of Na metal at $\hbar\omega=80$ eV for $\Theta =$ 0{\textdegree} and 90{\textdegree}.
Because of $\beta_{\nu n \rm s}^{\hbar\omega}=2$ for the $s$ orbitals, the contribution of the $3s$ orbital becomes zero
for $\Theta = 90${\textdegree}, as shown in Fig. \ref{fig_pes80eV_Na}(b).
Thus the $3p$ contribution to the valence band can be extracted by the $\Theta=90${\textdegree} measurement.
Here we note that
we have chosen 80 eV as the photon energy because Table \ref{tbl_phicsbetaAc_Na} shows
that the atomic-subshell PI-CS of Na $3s$ and $3p$ orbitals are the same order of magnitude at 80 eV, while Na $3p$ becomes relatively very small in higher photon energies.
We, thus, emphasize that our computer program is also useful for finding such experimental conditions prior to an experiment.

\begin{figure}[htb!]
\includegraphics[width=8.6cm]{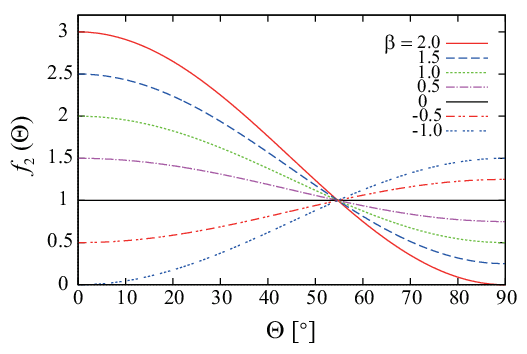}
\caption{\label{fig_f2} (Color online)
Dipole asymmetry factor $f_2(\Theta)=1+\beta P_2(\cos\Theta)$, which appears in Eq.~(\ref{eqn_pi-dcs}).
}
\end{figure}

In Fig.~\ref{fig_f2}, the $\Theta$ dependence of the dipole asymmetry factor,
$f_2(\Theta)=1+\beta P_2(\cos\Theta)$ in Eq.~(\ref{eqn_pi-dcs}),
is shown for several dipole coefficients $\beta$.
It is a monotonically decreasing function for $\beta>0$ while
it turns to be an increasing function for $\beta<0$. 
This remarkable feature can be used for efficient investigation of the atomic-subshell character in PES spectra by choosing appropriate $\Theta$ because $\beta$ depends on the subshells.

Figure~\ref{fig_f2} also demonstrates that the angle $\Theta_0=$ 54.7{\textdegree} is a special angle
for $f_2(\Theta)$ 
where $f_2(\Theta_0)$ becomes 1 and does not depend on $\beta$ due to $P_2(\cos\Theta_0)=0$.
Using $\Theta_0$ in Eq.~(\ref{eqn_pi-dcs}),
Eq.~(\ref{eqn_PE-DCS}) turns to be
\begin{eqnarray}
	P^{\hbar\omega}(\epsilon,\Theta_0)
	&=&
	\frac{1}{4\pi} \sum_{\nu(k) n l}^{\rm [as1]}
	\sigma^{\hbar\omega}_{\nu n l}
	D^{\rm as1}_{\nu(k) n l}(\epsilon) ~,
\end{eqnarray}
meaning that the PE-DCS at the angle $\Theta_0$ with a factor of $4\pi$ gives the total photoelectron cross section (PE-CS).
Actually, by integrating Eq.~(\ref{eqn_PE-DCS})
in terms of the solid angle, we obtain
\begin{eqnarray}
	2\pi\! \int_{0}^{\pi} \!\!\!
	\sin\Theta\,\diffa{\Theta}\,
	P^{\hbar\omega}(\epsilon,\Theta)
	&=&
	\sum_{\nu n l}^{\rm [as1]}
	\sigma^{\hbar\omega}_{\nu n l}
	D^{\rm as1}_{\nu(k) n l}(\epsilon) \\
	&=&
	4\pi P^{\hbar\omega}(\epsilon,\Theta_0)~.
\end{eqnarray}
Thus, the solid-angle integrated PES can be estimated
by a PES measurement at 
the ``magic" angle of
$\Theta=54.7^{\circ}$ for a polycrystalline sample.

\section{Application to a Delafossite Oxide\label{sec:CuAlO2r}}

\begin{figure}[htb!] 
\includegraphics[width=5cm]{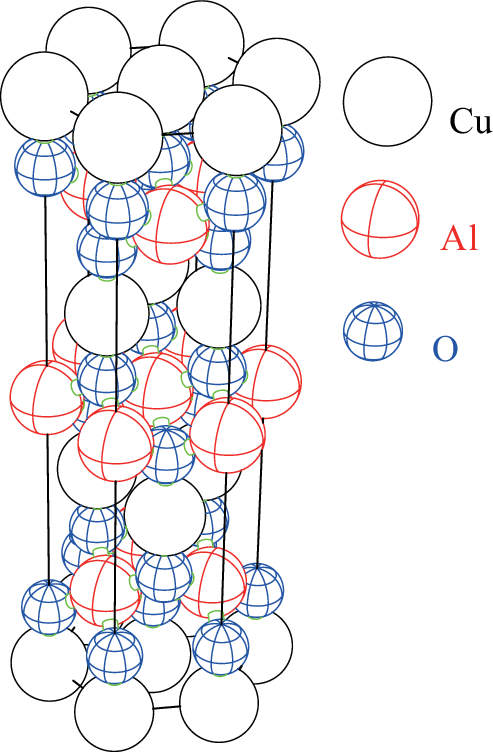}
\caption{\label{fig_cry_CuAlO2r} (Color online)
Crystal structure of a delafossite oxide CuAlO$_2$. Spheres are MTS's.
The nearest-neighbor distances are 3.51$a_0$ and 3.61$a_0$
for Cu-O and Al-O atomic pairs, respectively.
}
\end{figure}

Our method is applied to CuAlO$_2$ in the delafossite structure
shown in Fig. \ref{fig_cry_CuAlO2r}. The spase group is (\textit{R$\bar{3}$m}, 166), the rhombic lattice parameters are
$a=$ 2.8567 {\AA} and $c=$ 16.943 {\AA}, and
the atomic positions are $(0,0,0)$, $(0,0,1/2)$ and $(0,0,0.10977)$ for Cu, Al and O, respectively \cite{ishiguro83}.
The coordination numbers of Cu and Al are 2 and 6, respectively.

The band structure calculation has been performed
by using the code ABCAP \cite{hamada84a,yanase99,umekawa04}
with the full potential linearized augmented plane-wave (FLAPW) method \cite{andersen75,takeda79} in the LDA+$U$ scheme \cite{hohenberg64,kohn65,vosko80,painter81,anisimov91,solovyev96a,anisimov97}.
The effective Coulomb repulsion $U_{\rm eff}=U-J$ is 2.0 eV
for Cu $3d$ with $J=$0.7 eV. The plane-wave cut-off energy is 218 eV for the wave function. 
We take 1728 $\mib{k}$ points in the rhombohedral Brillouin zone.

\subsection{AS1 radius and PDOS}

Following the general procedure described in Sec.~\ref{ssec:PDOS},
$r_{\rm Cu}^{\rm as1}$ and $r_{\rm Al}^{\rm as1}$ have been determined by using the fcc Cu and Al metals and $r_{\rm O}^{\rm as1}$ has been determined to be a typical average of several oxide compounds.
As a result, the sum of the orbital-occupation numbers for each atomic element
$N_{\rm Cu}^{\rm as1}$
+$N_{\rm Al}^{\rm as1}$
+$N_{\rm O}^{\rm as1}$ 
approximately becomes the total valence-electron number $N_{\rm Cu}$+$N_{\rm Al}$+$N_{\rm O}$.
All the AS1 parameters are listed in Table \ref{tbl_f_CuAlO2r}.
Table~\ref{tbl_f_CuAlO2r} shows that the MTS-to-AS1 enhancement factors $f_{\nu n l}$ are small for Cu $3d$, moderate for O $2s,2p$, and large for Cu $4s,4p$ and Al $3s,3p$, depending on the degree of orbital localization.

\begin{table*}[htb!]
\caption{\label{tbl_f_CuAlO2r}
MTS radius $r_{\nu}^{\rm mts}$,
AS1 radius $r_{\nu}^{\rm as1}$,
MTS-to-AS1 enhancement factor $f_{\nu nl}$,
MTS orbital occupation number $N_{\nu nl}^{\rm mts}$,
and
AS1 orbital occupation number $N_{\nu nl}^{\rm as1}$
for $\nu=$ Cu, Al and O in CuAlO$_2$ crystal.
}
\begin{ruledtabular}
\begin{tabular}{|c|ccccc|cccc|cccc|}
$\nu$ &
\multicolumn{5}{c|}{Cu} & 
\multicolumn{4}{c|}{Al} & 
\multicolumn{4}{c|}{O} \\
$r_{\nu}^{\rm mts}/a_0$ & 
\multicolumn{5}{c|}{2.0} & 
\multicolumn{4}{c|}{1.8} & 
\multicolumn{4}{c|}{1.4} \\
$r_{\nu}^{\rm as1}/a_0$ & 
\multicolumn{5}{c|}{2.9} & 
\multicolumn{4}{c|}{3.6} & 
\multicolumn{4}{c|}{2.4} \\
$nl$ & $3d$ & $4s$ & $4p$ & $N_{\rm Cu}^{\rm as1}$ & $N_{\rm Cu}$ & 
$3s$ & $3p$ & $N_{\rm Al}^{\rm as1}$ & $N_{\rm Al}$ & 
$2s$ & $2p$ & $N_{\rm O}^{\rm as1}$ & $N_{\rm O}$ \\ \hline
$f_{\nu nl}$ & 
1.06 & 2.47 & 3.81 &  &  & 
3.48 & 5.86 &  &  & 
1.31 & 1.40 &  &  \\ 
$N_{\nu nl}^{\rm mts}$ & 
8.97 & 0.27 & 0.17 & (9.41) &  & 
0.24 & 0.30 & (0.54) &  & 
1.37 & 3.17 & 4.54 &  \\
$N_{\nu nl}^{\rm as1}$ & 
9.54 & 0.67 & 0.65 & 10.86 & 11 & 
0.82 & 1.74 & 2.56 & 3 & 
1.79 & 4.44 & 6.23 & 6
\end{tabular}
\end{ruledtabular}
\end{table*}

\begin{figure*}[htb!]
\centering
\includegraphics[width=14cm]{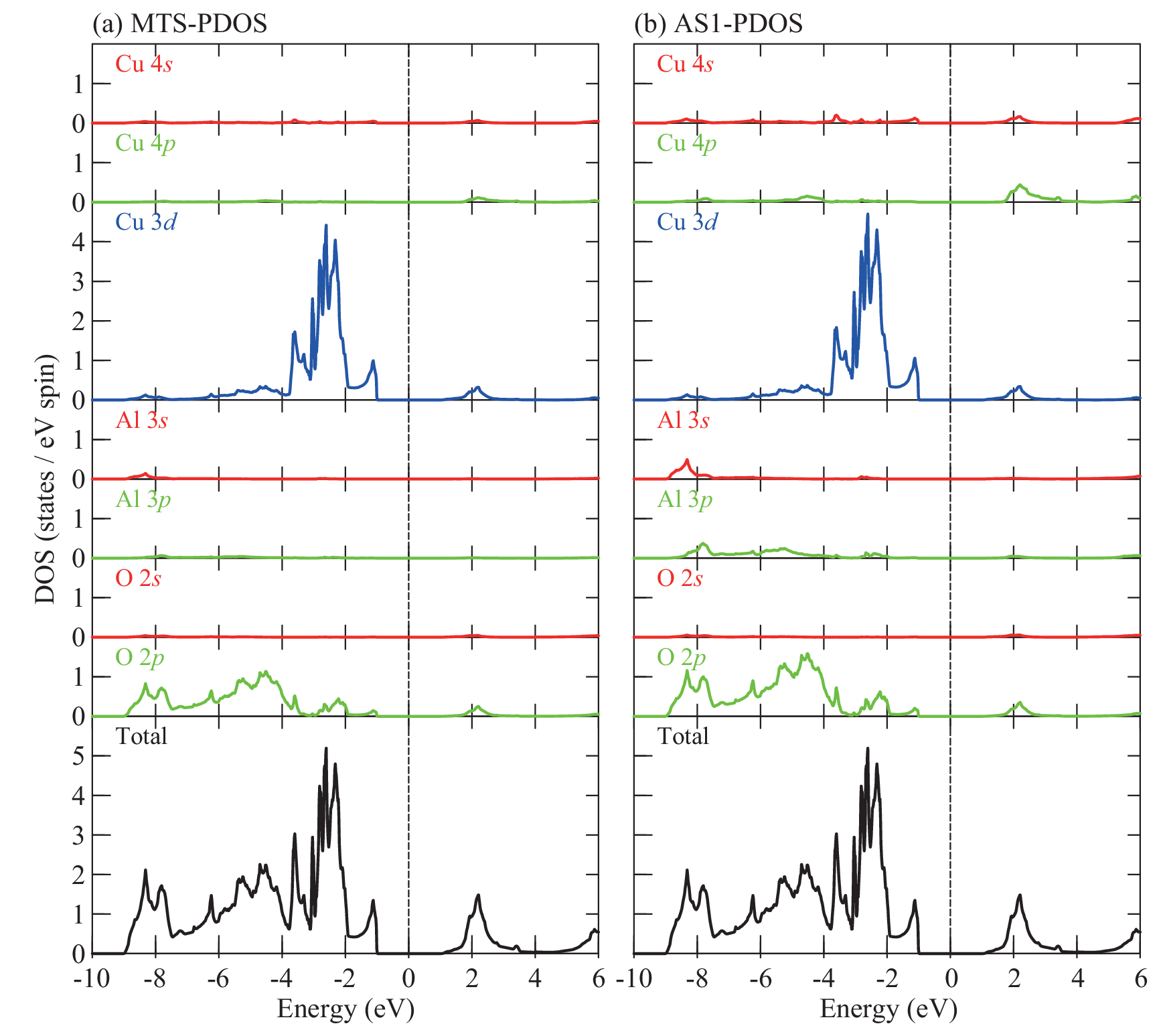}
\caption{\label{fig_PDOS_CuAlO2r} (Color online)
PDOS of Cu $4s,4p,3d$, Al $3s, 3p$, O $2s, 2p$, and the total DOS of CuAlO$_2$ calculated by using (a) MTS and (b) AS1.
Note that the both total DOS's are identical.
}
\end{figure*}

Figure~\ref{fig_PDOS_CuAlO2r} shows a comparison of the PDOS calculations on CuAlO$_2$ by using (a) MTS and (b) AS1, together with the total DOS.
The AS1-PDOS clearly shows which orbital actually contributes in the Cu-O and Al-O covalent bonds;
in the AS1-PDOS, the peak around 2 eV is due to the hybridization between the Cu $4s$$-$$4p$$-$$3d$ and the O $2p$ orbitals, while in the MTS-DOS, only the Cu $3d$$-$O $2p$ hybridization seems to be important.
One can also see that the Cu $3d$ orbital is slightly ionized because of the Cu $3d$ PDOS around 2 eV.
It is obvious in the AS1-PDOS that the peaks around $-$8 eV is due to the hybridization between the Al $3s$$-$$3p$ and the O $2p$ orbitals while in the MTS-PDOS, the O $2p$ states alone seem to contribute to the peak.

The nearest-neighbor distance of an A-B atomic pair $d_{\rm AB}$ is much smaller than the sum of those AS1 radii:
\begin{eqnarray*}
d_{\rm CuO}=3.51a_0 &\ll& 5.3a_0 =r_{\rm Cu}^{\rm as1}+r_{\rm O}^{\rm as1}\\
d_{\rm AlO}=3.61a_0 &\ll& 6.0a_0 =r_{\rm Al}^{\rm as1}+r_{\rm O}^{\rm as1}
\end{eqnarray*}
Thus, the AS1 atomic spheres considerably overlap with each other in the crystal.
It is worthwhile to stress that the Al $3d$ orbital is not included in the AS1 orbital, even though the MTS-PDOS has the $3d$ component. This is because the $3d$ component is considered to be made of the nearest-neighbor O $2p$ orbitals.

We expect that the same values of $r_{\nu}^{\rm as1}$'s can be applied to other materials, although we have to wait for a future study on that issue.

\subsection{Atomic-subshell PI-DCS}

Table \ref{tbl_phicsbeta_CuAlO2r} shows the atomic-subshell PI-CS and the dipole coefficient of Cu, Al, and O calculated at several photon energies.
The PI-CS is the largest for Cu $3d$ and the second largest for O $2p$ among these orbitals at $\hbar\omega=80$ eV, while it is large for all the $s$ orbitals and small for the other orbitals at $\hbar\omega=7940$ eV.
All the PI-CS's rapidly decrease with increasing photon energy. The decreasing rate is smaller in the $s$ orbitals than in the other orbitals.
Thus, the PES intensity from the $s$ orbitals becomes dominant in the hard x-ray region.

\begin{table}[htb!] 
\caption{\label{tbl_phicsbeta_CuAlO2r}
Atomic-subshell PI-CS per electron $\sigma_{\nu n l}^{\hbar\omega}$ and the dipole coefficient $\beta_{\nu n l}^{\hbar\omega}$ for each atomic orbital in CuAlO$_2$
calculated for several photo energies.
} 
\begin{ruledtabular}
\begin{tabular}{|c|ccc|cc|cc|}
$\nu$ & 
\multicolumn{3}{c|}{Cu} & 
\multicolumn{2}{c|}{Al} & 
\multicolumn{2}{c|}{O}  \\ 
$nl$ & 
$3d$ & $4s$ & $4p$ & $3s$ & $3p$ & $2s$ & $2p$ \\ \hline
$\hbar\omega =$ 80 eV 
&  &  &  &  &  &  &  \\
$\sigma_{\nu nl}^{\hbar\omega}$ [kb] &
902.5 & {27.7} & 3.7 & {77.7} & 66.1 & {346.1} & 515.0 \\
$\beta_{\nu nl}^{\hbar\omega}$ &
0.51 & {2.00}& 0.28 & {2.00} & 1.50 & {2.00} & 1.50 \\
\hline
$\hbar\omega =$ 320 eV
&  &  &  &  &  &  &  \\
$\sigma_{\nu nl}^{\hbar\omega}$ [kb] & 
96.00& {4.32}& 1.37 & {8.47} & 3.87 & {36.87} & 11.13 \\
$\beta_{\nu nl}^{\hbar\omega}$ & 
1.17 & {2.00}& 1.60 & {2.00} & 1.51 & {2.00} & 1.20 \\
\hline
$\hbar\omega =$ 1500 eV
&  &  &  &  &  &  &  \\
$\sigma_{\nu nl}^{\hbar\omega}$ [b] & 
798.4& {306.7} & 68.4 & {356.4} & 43.9 & {934.3}& 63.5 \\
$\beta_{\nu nl}^{\hbar\omega}$ & 
0.95 & {2.00} & 1.51 & {2.00} & 0.92 & {2.00} & 0.51 \\  
\hline
$\hbar\omega =$ 7940 eV
&  &  &  &  &  &  &  \\
$\sigma_{\nu nl}^{\hbar\omega}$ [b] & 
1.09 & {8.52} & 0.57 & {4.64} & 0.12 & {7.78} & 0.10  \\
$\beta_{\nu nl}^{\hbar\omega}$ & 
0.42 & {2.00} & 0.88 & {2.00} & 0.30 & {2.00} & 0.12  \end{tabular}
\end{ruledtabular}
\end{table}

\subsection{PE-DCS -- HX-PES --}

\begin{figure}[htb!]
\centering
\includegraphics[width=7cm]{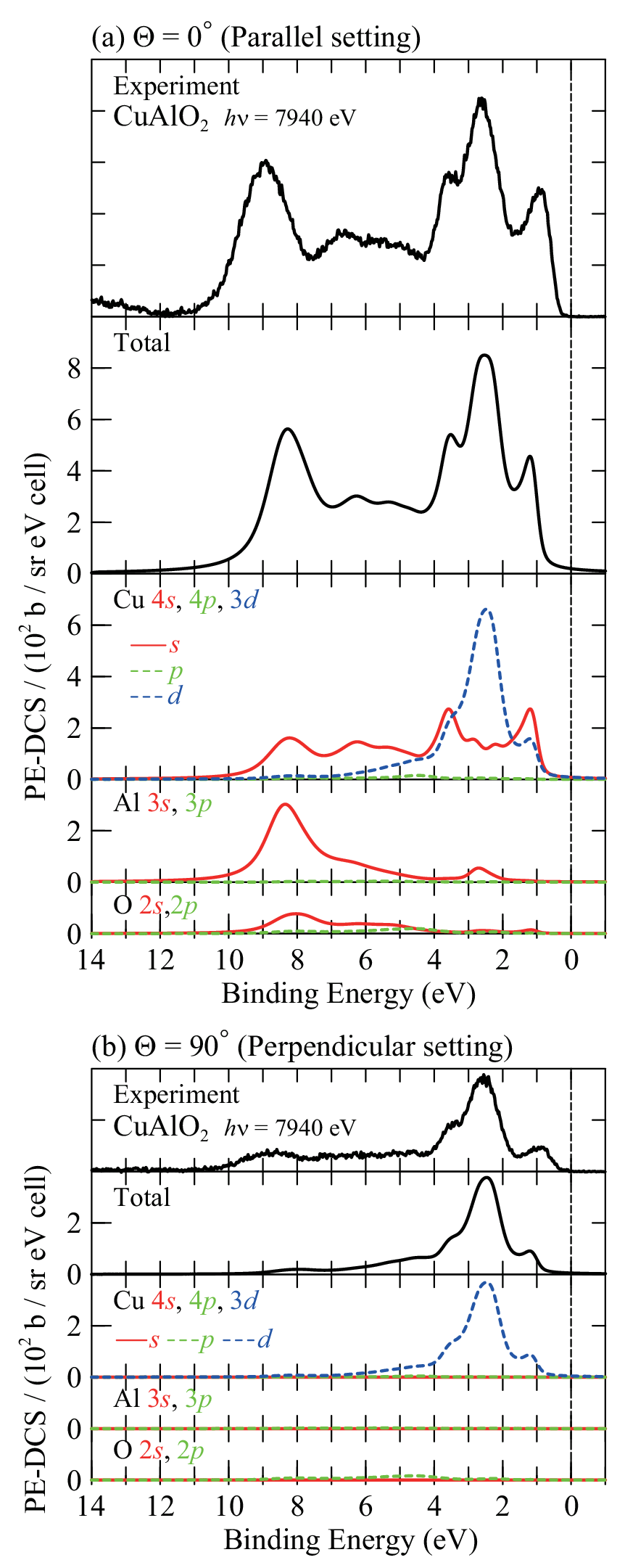}
\caption{\label{fig_7940eV_CuAlO2r} (Color online)
PE-DCS of CuAlO$_2$ at $\hbar\omega=$7940 eV for (a) $\Theta = 0${\textdegree} and (b) $\Theta = 90${\textdegree}.
The vertical axis of the experiment is in arbitrary units.
In the calculation, the experimental resolution is set to be 250 meV and the Lorentzian broadening due to life time effect is also considered \cite{takahashi19,Note4}.
}
\end{figure}

Figure~\ref{fig_7940eV_CuAlO2r} shows a comparison between the PE-DCS of CuAlO$_2$ at $\hbar\omega=$7940 eV and the experimental HX-PES spectra \cite{takahashi19}.
It is very obvious that the theory well reproduces all the characteristic features of the experiment while the conventional MTS-PDOS analysis have failed to reproduce the experiment \cite{takahashi19}.

In the parallel ($\Theta=$ 0{\textdegree}) setting, the major contributions to PE-DCS are from all the $s$ orbitals and the Cu $3d$ orbital.
At the valence band bottom, the O $2p$ orbital is strongly hybridized with the Al $3s, 3p$ orbitals, rather than the Cu $3d$ orbital (see Fig. \ref{fig_PDOS_CuAlO2r}),
which should be observed in the spectrum via the Cu $4s$, Al $3s$, and O $2s$ spectral weight due to their large PI-CS's (see TABLE~\ref{tbl_phicsbeta_CuAlO2r}).
This theoretical prediction is actually confirmed by the large 9-eV peak in the experimental spectrum.

A combination of (1) the large $s$-O $2p$ hybridization, which is severely underestimated in the conventional MTS-PDOS analysis, and (2) the very large $s$-orbital contribution in the hard x-ray region, is the reason why the conventional MTS-PDOS analysis fails to reproduce HX-PES spectra, not only of CuAlO$_2$ but of many transition-metal (TM) oxides; 
in the conventional MTS-PDOS analysis on TM oxides, the larger O $2p$ weight has often been taken into account empirically \cite{Saitoh97}. However, the whole TM $4s$ and O $2s$ PDOS have always been ignored because of their small DOS's. In the low photon energy region, this analysis seems to have been good owing to their small PI-CS, thereby the problem has been hidden.
A similar situation is recently observed in La $5p$ spectral weight in the valence band HX-PES spectra of lanthanum compounds such as LaCoO$_3$ \cite{Takegami19} or La$_{1-x}$Sr$_x$MnO$_3$ \cite{Note5}.

The agreement between theory and experiment in the perpenducular ($\Theta=$ 90{\textdegree}) setting is still good, but looks a little worse than the parallel setting.
Here, the contributions of all the $s$ orbitals to the PI-DCS disappear, and only the Cu $3d$ has a major contribution.
Therefore, 
this may be reflecting that the calculations on the Cu $3d$ state is poor in comparison with the $s$ and $p$ states;
the LDA+$U$ calculation cannot appropriately take into account 
the Cu $3d^9$ final state, or the satellite structures in the PES.
Although the effect exists also in the parallel setting, it is hidden by the strong s orbital intensity.

\section{Conclusion\label{sec:conclusion}}

We have developed a method to calculate the photoelectron differential cross section (PE-DCS) of the valence band for the linearly polarized photon from the partial density of states (PDOS) and the atomic-orbital photoionization differential cross section (PI-DCS).

Beyond the MTS-PDOS evaluated in the muffin-tin sphere (MTS), we have proposed a new AS1-PDOS as an appropriate PDOS, and have shown that the AS1-PDOS is nicely evaluated
by multiplying an atomic-orbital-dependent enhancement factor ($f_{\nu nl}$).

By employing the Hartree-Fock-Slater method, we have developed a new computer code to calculate the PI-DCS of all the needed atomic orbitals including the orbitals unoccupied in the ground-state atom for the incident photon energy up to $10^4$ eV.

Details of the PE-DCS calculation have been examined by using simple examples of Na metal and NaCl.
In application to CuAlO$_2$, we have shown the strong ability of our method for HX-PES.
However, further study is still needed for the photon-polarization dependence on the PE-DCS.
For example, the selection rule in the spherical atom may be
too strong for the crystals with the lower symmetry.
Overall features of the atomic-limit model are satisfactory, and the essential part of the problems in reproducing HX-PES spectra raised in the previous paper \cite{takahashi19} is well solved.

\begin{acknowledgments}
This work was supported by JSPS KAKENHI Grant Numbers 17K05502, 18K03503, 18K03543, 18K03550, 21H01536, and 21K03474.
The synchrotron radiation experiments were performed at the BL09XU and BL47XU of SPring-8 with the approval of the Japan Synchrotron Radiation Research Institute (JASRI) (Proposal Nos. 
2014B1011, 2015A1008, 2016A1013, 2016B1006, 2017A1250, 2017B1026, 2017B1303, 2018A1013, 2018B1025, 2019B1013, 2020A1008, 2020A1258).
\end{acknowledgments}


\begin{thebibliography}{35}%
	\makeatletter
	\providecommand \@ifxundefined [1]{%
		\@ifx{#1\undefined}
	}%
	\providecommand \@ifnum [1]{%
		\ifnum #1\expandafter \@firstoftwo
		\else \expandafter \@secondoftwo
		\fi
	}%
	\providecommand \@ifx [1]{%
		\ifx #1\expandafter \@firstoftwo
		\else \expandafter \@secondoftwo
		\fi
	}%
	\providecommand \natexlab [1]{#1}%
	\providecommand \enquote  [1]{``#1''}%
	\providecommand \bibnamefont  [1]{#1}%
	\providecommand \bibfnamefont [1]{#1}%
	\providecommand \citenamefont [1]{#1}%
	\providecommand \href@noop [0]{\@secondoftwo}%
	\providecommand \href [0]{\begingroup \@sanitize@url \@href}%
	\providecommand \@href[1]{\@@startlink{#1}\@@href}%
	\providecommand \@@href[1]{\endgroup#1\@@endlink}%
	\providecommand \@sanitize@url [0]{\catcode `\\12\catcode `\$12\catcode
		`\&12\catcode `\#12\catcode `\^12\catcode `\_12\catcode `\%12\relax}%
	\providecommand \@@startlink[1]{}%
	\providecommand \@@endlink[0]{}%
	\providecommand \url  [0]{\begingroup\@sanitize@url \@url }%
	\providecommand \@url [1]{\endgroup\@href {#1}{\urlprefix }}%
	\providecommand \urlprefix  [0]{URL }%
	\providecommand \Eprint [0]{\href }%
	\providecommand \doibase [0]{https://doi.org/}%
	\providecommand \selectlanguage [0]{\@gobble}%
	\providecommand \bibinfo  [0]{\@secondoftwo}%
	\providecommand \bibfield  [0]{\@secondoftwo}%
	\providecommand \translation [1]{[#1]}%
	\providecommand \BibitemOpen [0]{}%
	\providecommand \bibitemStop [0]{}%
	\providecommand \bibitemNoStop [0]{.\EOS\space}%
	\providecommand \EOS [0]{\spacefactor3000\relax}%
	\providecommand \BibitemShut  [1]{\csname bibitem#1\endcsname}%
	\let\auto@bib@innerbib\@empty
	\bibitem [{Note2()}]{Note2}%
	\BibitemOpen
	\bibinfo {note} {For simplicity, we hereafter describe valence-band
		angle-integrated PES spectra as valence-band PES spectra.}\BibitemShut
	{Stop}%
	\bibitem [{\citenamefont {Yeh}\ and\ \citenamefont {Lindau}(1985)}]{yeh85}%
	\BibitemOpen
	\bibfield  {author} {\bibinfo {author} {\bibfnamefont {J.~J.}\ \bibnamefont
			{Yeh}}\ and\ \bibinfo {author} {\bibfnamefont {I.}~\bibnamefont {Lindau}},\
	}\href@noop {} {\bibfield  {journal} {\bibinfo  {journal} {At. Data Nucl.
				Data Tables}\ }\textbf {\bibinfo {volume} {32}},\ \bibinfo {pages} {1}
		(\bibinfo {year} {1985})}\BibitemShut {NoStop}%
	\bibitem [{\citenamefont {Trzhaskovskaya}\ \emph {et~al.}(2001)\citenamefont
		{Trzhaskovskaya}, \citenamefont {Nefedov},\ and\ \citenamefont
		{Yarzhemsky}}]{trzhaskovskaya01}%
	\BibitemOpen
	\bibfield  {author} {\bibinfo {author} {\bibfnamefont {M.~B.}\ \bibnamefont
			{Trzhaskovskaya}}, \bibinfo {author} {\bibfnamefont {V.~I.}\ \bibnamefont
			{Nefedov}},\ and\ \bibinfo {author} {\bibfnamefont {V.~G.}\ \bibnamefont
			{Yarzhemsky}},\ }\href@noop {} {\bibfield  {journal} {\bibinfo  {journal}
			{At. Data Nucl. Data Tables}\ }\textbf {\bibinfo {volume} {77}},\ \bibinfo
		{pages} {97} (\bibinfo {year} {2001})}\BibitemShut {NoStop}%
	\bibitem [{\citenamefont {Trzhaskovskaya}\ \emph {et~al.}(2002)\citenamefont
		{Trzhaskovskaya}, \citenamefont {Nefedov},\ and\ \citenamefont
		{Yarzhemsky}}]{trzhaskovskaya02}%
	\BibitemOpen
	\bibfield  {author} {\bibinfo {author} {\bibfnamefont {M.~B.}\ \bibnamefont
			{Trzhaskovskaya}}, \bibinfo {author} {\bibfnamefont {V.~I.}\ \bibnamefont
			{Nefedov}},\ and\ \bibinfo {author} {\bibfnamefont {V.~G.}\ \bibnamefont
			{Yarzhemsky}},\ }\href@noop {} {\bibfield  {journal} {\bibinfo  {journal}
			{At. Data Nucl. Data Tables}\ }\textbf {\bibinfo {volume} {82}},\ \bibinfo
		{pages} {257} (\bibinfo {year} {2002})}\BibitemShut {NoStop}%
	\bibitem [{Note3()}]{Note3}%
	\BibitemOpen
	\bibinfo {note} {We note that PE-DCS includes all the photoelectrons to the
		all directions of the crystallographic orientation because we are assuming
		polycrystalline samples and/or rough surfaces and the crystallographic
		anisotropy will be averaged out. Therefore, PE-DCS is not equal to the
		angle-resolved PES (ARPES) intensity from a single crystalline
		sample.}\BibitemShut {Stop}%
	\bibitem [{\citenamefont {Saitoh}\ \emph {et~al.}(1997)\citenamefont {Saitoh},
		\citenamefont {Mizokawa}, \citenamefont {Fujimori}, \citenamefont {Abbate},
		\citenamefont {Takeda},\ and\ \citenamefont {Takano}}]{Saitoh97}%
	\BibitemOpen
	\bibfield  {author} {\bibinfo {author} {\bibfnamefont {T.}~\bibnamefont
			{Saitoh}}, \bibinfo {author} {\bibfnamefont {T.}~\bibnamefont {Mizokawa}},
		\bibinfo {author} {\bibfnamefont {A.}~\bibnamefont {Fujimori}}, \bibinfo
		{author} {\bibfnamefont {M.}~\bibnamefont {Abbate}}, \bibinfo {author}
		{\bibfnamefont {Y.}~\bibnamefont {Takeda}},\ and\ \bibinfo {author}
		{\bibfnamefont {M.}~\bibnamefont {Takano}},\ }\href@noop {} {\bibfield
		{journal} {\bibinfo  {journal} {Phys.\ Rev. B}\ }\textbf {\bibinfo {volume}
			{55}},\ \bibinfo {pages} {4257} (\bibinfo {year} {1997})}\BibitemShut
	{NoStop}%
	\bibitem [{\citenamefont {Trzhaskovskaya}\ \emph {et~al.}(2018)\citenamefont
		{Trzhaskovskaya}, \citenamefont {Nefedov},\ and\ \citenamefont
		{Yarzhemsky}}]{trzhaskovskaya18}%
	\BibitemOpen
	\bibfield  {author} {\bibinfo {author} {\bibfnamefont {M.~B.}\ \bibnamefont
			{Trzhaskovskaya}}, \bibinfo {author} {\bibfnamefont {V.~I.}\ \bibnamefont
			{Nefedov}},\ and\ \bibinfo {author} {\bibfnamefont {V.~G.}\ \bibnamefont
			{Yarzhemsky}},\ }\href@noop {} {\bibfield  {journal} {\bibinfo  {journal}
			{At. Data Nucl. Data Tables}\ }\textbf {\bibinfo {volume} {119}},\ \bibinfo
		{pages} {99} (\bibinfo {year} {2018})}\BibitemShut {NoStop}%
	\bibitem [{\citenamefont {Bagheri}\ and\ \citenamefont
		{Blaha}(2019)}]{Bagheri19}%
	\BibitemOpen
	\bibfield  {author} {\bibinfo {author} {\bibfnamefont {M.}~\bibnamefont
			{Bagheri}}\ and\ \bibinfo {author} {\bibfnamefont {P.}~\bibnamefont
			{Blaha}},\ }\href@noop {} {\bibfield  {journal} {\bibinfo  {journal} {J.
				Electron Spectrsc. Relat. Phenom.}\ }\textbf {\bibinfo {volume} {230}},\
		\bibinfo {pages} {1} (\bibinfo {year} {2019})}\BibitemShut {NoStop}%
	\bibitem [{\citenamefont {Blaha}\ \emph {et~al.}(2020)\citenamefont {Blaha},
		\citenamefont {Schwarz}, \citenamefont {Tran}, \citenamefont {Laskowski},
		\citenamefont {Madsen},\ and\ \citenamefont {Marks}}]{Blaha20}%
	\BibitemOpen
	\bibfield  {author} {\bibinfo {author} {\bibfnamefont {P.}~\bibnamefont
			{Blaha}}, \bibinfo {author} {\bibfnamefont {K.}~\bibnamefont {Schwarz}},
		\bibinfo {author} {\bibfnamefont {F.}~\bibnamefont {Tran}}, \bibinfo {author}
		{\bibfnamefont {R.}~\bibnamefont {Laskowski}}, \bibinfo {author}
		{\bibfnamefont {G.~K.~H.}\ \bibnamefont {Madsen}},\ and\ \bibinfo {author}
		{\bibfnamefont {L.~D.}\ \bibnamefont {Marks}},\ }\href@noop {} {\bibfield
		{journal} {\bibinfo  {journal} {J. Chem. Phys.}\ }\textbf {\bibinfo {volume}
			{152}},\ \bibinfo {pages} {074101} (\bibinfo {year} {2020})}\BibitemShut
	{NoStop}%
	\bibitem [{\citenamefont {Takahashi}\ \emph {et~al.}(2019)\citenamefont
		{Takahashi}, \citenamefont {Kato}, \citenamefont {Okawa}, \citenamefont
		{Okuda}, \citenamefont {Yasui}, \citenamefont {Ikenaga}, \citenamefont {Ono},
		\citenamefont {Hamada},\ and\ \citenamefont {Saitoh}}]{takahashi19}%
	\BibitemOpen
	\bibfield  {author} {\bibinfo {author} {\bibfnamefont {K.}~\bibnamefont
			{Takahashi}}, \bibinfo {author} {\bibfnamefont {R.}~\bibnamefont {Kato}},
		\bibinfo {author} {\bibfnamefont {M.}~\bibnamefont {Okawa}}, \bibinfo
		{author} {\bibfnamefont {T.}~\bibnamefont {Okuda}}, \bibinfo {author}
		{\bibfnamefont {A.}~\bibnamefont {Yasui}}, \bibinfo {author} {\bibfnamefont
			{E.}~\bibnamefont {Ikenaga}}, \bibinfo {author} {\bibfnamefont
			{K.}~\bibnamefont {Ono}}, \bibinfo {author} {\bibfnamefont {N.}~\bibnamefont
			{Hamada}},\ and\ \bibinfo {author} {\bibfnamefont {T.}~\bibnamefont
			{Saitoh}},\ }\href@noop {} {\bibfield  {journal} {\bibinfo  {journal} {J.
				Phys. Soc. Jpn}\ }\textbf {\bibinfo {volume} {88}},\ \bibinfo {pages} {074701
		} (\bibinfo {year} {2019})}\BibitemShut {NoStop}%
	\bibitem [{\citenamefont {Weinen}\ \emph {et~al.}(2015)\citenamefont {Weinen},
		\citenamefont {Koethe}, \citenamefont {Chang}, \citenamefont {Agrestini},
		\citenamefont {Kasinathan}, \citenamefont {Liao}, \citenamefont {Fujiwara},
		\citenamefont {Sch{\"u}{\ss}ler-Langeheine}, \citenamefont {Strigari},
		\citenamefont {Haupricht}, \citenamefont {Panaccione}, \citenamefont {Offi},
		\citenamefont {Monaco}, \citenamefont {Huotari}, \citenamefont {Tsuei},\ and\
		\citenamefont {Tjeng}}]{Weinen15}%
	\BibitemOpen
	\bibfield  {author} {\bibinfo {author} {\bibfnamefont {J.}~\bibnamefont
			{Weinen}}, \bibinfo {author} {\bibfnamefont {T.~C.}\ \bibnamefont {Koethe}},
		\bibinfo {author} {\bibfnamefont {C.~F.}\ \bibnamefont {Chang}}, \bibinfo
		{author} {\bibfnamefont {S.}~\bibnamefont {Agrestini}}, \bibinfo {author}
		{\bibfnamefont {D.}~\bibnamefont {Kasinathan}}, \bibinfo {author}
		{\bibfnamefont {Y.~F.}\ \bibnamefont {Liao}}, \bibinfo {author}
		{\bibfnamefont {H.}~\bibnamefont {Fujiwara}}, \bibinfo {author}
		{\bibfnamefont {C.}~\bibnamefont {Sch{\"u}{\ss}ler-Langeheine}}, \bibinfo
		{author} {\bibfnamefont {F.}~\bibnamefont {Strigari}}, \bibinfo {author}
		{\bibfnamefont {T.}~\bibnamefont {Haupricht}}, \bibinfo {author}
		{\bibfnamefont {G.}~\bibnamefont {Panaccione}}, \bibinfo {author}
		{\bibfnamefont {F.}~\bibnamefont {Offi}}, \bibinfo {author} {\bibfnamefont
			{G.}~\bibnamefont {Monaco}}, \bibinfo {author} {\bibfnamefont
			{S.}~\bibnamefont {Huotari}}, \bibinfo {author} {\bibfnamefont {K.-D.}\
			\bibnamefont {Tsuei}},\ and\ \bibinfo {author} {\bibfnamefont {L.~H.}\
			\bibnamefont {Tjeng}},\ }\href@noop {} {\bibfield  {journal} {\bibinfo
			{journal} {J. Electron Spectrosc. Relat. Phenom.}\ }\textbf {\bibinfo
			{volume} {198}},\ \bibinfo {pages} {6} (\bibinfo {year} {2015})}\BibitemShut
	{NoStop}%
	\bibitem [{\citenamefont {Kobayashi}(2009)}]{Kobayashi09}%
	\BibitemOpen
	\bibfield  {author} {\bibinfo {author} {\bibfnamefont {K.}~\bibnamefont
			{Kobayashi}},\ }\href@noop {} {\bibfield  {journal} {\bibinfo  {journal}
			{Nucl. Instrum. Methods. Phys. Res. A}\ }\textbf {\bibinfo {volume} {601}},\
		\bibinfo {pages} {32} (\bibinfo {year} {2009})}\BibitemShut {NoStop}%
	\bibitem [{Note6()}]{Note6}%
	\BibitemOpen
	\bibinfo {note} {Because we define $r^{\rm as1}$ as a ``nearly-neutral atomic
		sphere'' $N_{\nu(k)}^{\rm as1}$ is ideally equal to the valence-electron
		number of an isolated atom of atomic element $\nu$, which is common to all
		the inequivalent sites $k$. Therefore, we define $N_{\nu}$ without
		$(k)$.}\BibitemShut {Stop}%
	\bibitem [{\citenamefont {Inoshita}\ \emph {et~al.}(2014)\citenamefont
		{Inoshita}, \citenamefont {Jeong}, \citenamefont {Hamada},\ and\
		\citenamefont {Hosono}}]{inoshita14}%
	\BibitemOpen
	\bibfield  {author} {\bibinfo {author} {\bibfnamefont {T.}~\bibnamefont
			{Inoshita}}, \bibinfo {author} {\bibfnamefont {S.}~\bibnamefont {Jeong}},
		\bibinfo {author} {\bibfnamefont {N.}~\bibnamefont {Hamada}},\ and\ \bibinfo
		{author} {\bibfnamefont {H.}~\bibnamefont {Hosono}},\ }\href@noop {}
	{\bibfield  {journal} {\bibinfo  {journal} {Phys. Rev. X}\ }\textbf {\bibinfo
			{volume} {4}},\ \bibinfo {pages} {031023} (\bibinfo {year}
		{2014})}\BibitemShut {NoStop}%
	\bibitem [{\citenamefont {Inoshita}\ \emph {et~al.}(2015)\citenamefont
		{Inoshita}, \citenamefont {Hamada},\ and\ \citenamefont
		{Hosono}}]{inoshita15}%
	\BibitemOpen
	\bibfield  {author} {\bibinfo {author} {\bibfnamefont {T.}~\bibnamefont
			{Inoshita}}, \bibinfo {author} {\bibfnamefont {N.}~\bibnamefont {Hamada}},\
		and\ \bibinfo {author} {\bibfnamefont {H.}~\bibnamefont {Hosono}},\
	}\href@noop {} {\bibfield  {journal} {\bibinfo  {journal} {Phys. Rev. B}\
		}\textbf {\bibinfo {volume} {92}},\ \bibinfo {pages} {201109(R)} (\bibinfo
		{year} {2015})}\BibitemShut {NoStop}%
	\bibitem [{\citenamefont {Cooper}\ and\ \citenamefont {Zare}(1968)}]{cooper68}%
	\BibitemOpen
	\bibfield  {author} {\bibinfo {author} {\bibfnamefont {J.~W.}\ \bibnamefont
			{Cooper}}\ and\ \bibinfo {author} {\bibnamefont {Zare}},\ }\href@noop {}
	{\bibfield  {journal} {\bibinfo  {journal} {J. Chem. Phys.}\ }\textbf
		{\bibinfo {volume} {48}},\ \bibinfo {pages} {942} (\bibinfo {year}
		{1968})}\BibitemShut {NoStop}%
	\bibitem [{\citenamefont {Cooper}(1993)}]{cooper93}%
	\BibitemOpen
	\bibfield  {author} {\bibinfo {author} {\bibfnamefont {J.~W.}\ \bibnamefont
			{Cooper}},\ }\href@noop {} {\bibfield  {journal} {\bibinfo  {journal} {Phys.\
				Rev. A}\ }\textbf {\bibinfo {volume} {47}},\ \bibinfo {pages} {1841}
		(\bibinfo {year} {1993})}\BibitemShut {NoStop}%
	\bibitem [{\citenamefont {Scofield}(1973)}]{scofield73}%
	\BibitemOpen
	\bibfield  {author} {\bibinfo {author} {\bibfnamefont {J.~H.}\ \bibnamefont
			{Scofield}},\ }\href@noop {} {\bibfield  {journal} {\bibinfo  {journal}
			{\textit{Lawrence Livermore Laboratory Report,} UCRL-51326}\ } (\bibinfo
		{year} {1973})}\BibitemShut {NoStop}%
	\bibitem [{\citenamefont {Iwasawa}\ \emph {et~al.}(2009)\citenamefont
		{Iwasawa}, \citenamefont {Kaneyoshi}, \citenamefont {Kurahashi},
		\citenamefont {Saitoh}, \citenamefont {Hase}, \citenamefont {Katsufuji},
		\citenamefont {Shimada}, \citenamefont {Namatame},\ and\ \citenamefont
		{Taniguchi}}]{Iwasawa09}%
	\BibitemOpen
	\bibfield  {author} {\bibinfo {author} {\bibfnamefont {H.}~\bibnamefont
			{Iwasawa}}, \bibinfo {author} {\bibfnamefont {S.}~\bibnamefont {Kaneyoshi}},
		\bibinfo {author} {\bibfnamefont {K.}~\bibnamefont {Kurahashi}}, \bibinfo
		{author} {\bibfnamefont {T.}~\bibnamefont {Saitoh}}, \bibinfo {author}
		{\bibfnamefont {I.}~\bibnamefont {Hase}}, \bibinfo {author} {\bibfnamefont
			{T.}~\bibnamefont {Katsufuji}}, \bibinfo {author} {\bibfnamefont
			{K.}~\bibnamefont {Shimada}}, \bibinfo {author} {\bibfnamefont
			{H.}~\bibnamefont {Namatame}},\ and\ \bibinfo {author} {\bibfnamefont
			{M.}~\bibnamefont {Taniguchi}},\ }\href@noop {} {\bibfield  {journal}
		{\bibinfo  {journal} {Phys.\ Rev. B}\ }\textbf {\bibinfo {volume} {80}},\
		\bibinfo {pages} {125122} (\bibinfo {year} {2009})}\BibitemShut {NoStop}%
	\bibitem [{Note4()}]{Note4}%
	\BibitemOpen
	\bibinfo {note} {The binding energy ($E$) dependent Lorentzian FWHM due to
		lifetime effect is set to be $0.1(E-E_{\rm F})$ (eV).}\BibitemShut {Stop}%
	\bibitem [{\citenamefont {Ishiguro}\ \emph {et~al.}(1983)\citenamefont
		{Ishiguro}, \citenamefont {Ishizawa}, \citenamefont {Mizutani}, \citenamefont
		{Kato}, \citenamefont {Tanaka},\ and\ \citenamefont {Marumo}}]{ishiguro83}%
	\BibitemOpen
	\bibfield  {author} {\bibinfo {author} {\bibfnamefont {T.}~\bibnamefont
			{Ishiguro}}, \bibinfo {author} {\bibfnamefont {N.}~\bibnamefont {Ishizawa}},
		\bibinfo {author} {\bibfnamefont {N.}~\bibnamefont {Mizutani}}, \bibinfo
		{author} {\bibfnamefont {M.}~\bibnamefont {Kato}}, \bibinfo {author}
		{\bibfnamefont {K.}~\bibnamefont {Tanaka}},\ and\ \bibinfo {author}
		{\bibfnamefont {F.}~\bibnamefont {Marumo}},\ }\href@noop {} {\bibfield
		{journal} {\bibinfo  {journal} {Acta Cryst. B}\ }\textbf {\bibinfo {volume}
			{39}},\ \bibinfo {pages} {564} (\bibinfo {year} {1983})}\BibitemShut
	{NoStop}%
	\bibitem [{\citenamefont {Hamada}\ \emph {et~al.}(1984)\citenamefont {Hamada},
		\citenamefont {Terakura},\ and\ \citenamefont {Yanase}}]{hamada84a}%
	\BibitemOpen
	\bibfield  {author} {\bibinfo {author} {\bibfnamefont {N.}~\bibnamefont
			{Hamada}}, \bibinfo {author} {\bibfnamefont {K.}~\bibnamefont {Terakura}},\
		and\ \bibinfo {author} {\bibfnamefont {A.}~\bibnamefont {Yanase}},\
	}\href@noop {} {\bibfield  {journal} {\bibinfo  {journal} {J. Phys. F: Metal
				Phys.}\ }\textbf {\bibinfo {volume} {14}},\ \bibinfo {pages} {2371} (\bibinfo
		{year} {1984})}\BibitemShut {NoStop}%
	\bibitem [{\citenamefont {Yanase}\ and\ \citenamefont
		{Hamada}(1999)}]{yanase99}%
	\BibitemOpen
	\bibfield  {author} {\bibinfo {author} {\bibfnamefont {A.}~\bibnamefont
			{Yanase}}\ and\ \bibinfo {author} {\bibfnamefont {N.}~\bibnamefont
			{Hamada}},\ }\href@noop {} {\bibfield  {journal} {\bibinfo  {journal} {J.
				Phys. Soc. Jpn}\ }\textbf {\bibinfo {volume} {68}},\ \bibinfo {pages} {1607}
		(\bibinfo {year} {1999})}\BibitemShut {NoStop}%
	\bibitem [{\citenamefont {Umekawa}\ \emph {et~al.}(2004)\citenamefont
		{Umekawa}, \citenamefont {Hamada}, \citenamefont {Kodama},\ and\
		\citenamefont {Moritomo}}]{umekawa04}%
	\BibitemOpen
	\bibfield  {author} {\bibinfo {author} {\bibfnamefont {M.}~\bibnamefont
			{Umekawa}}, \bibinfo {author} {\bibfnamefont {N.}~\bibnamefont {Hamada}},
		\bibinfo {author} {\bibfnamefont {A.}~\bibnamefont {Kodama}},\ and\ \bibinfo
		{author} {\bibfnamefont {Y.}~\bibnamefont {Moritomo}},\ }\href@noop {}
	{\bibfield  {journal} {\bibinfo  {journal} {J. Phys. Soc. Jpn.}\ }\textbf
		{\bibinfo {volume} {72}},\ \bibinfo {pages} {430} (\bibinfo {year}
		{2004})}\BibitemShut {NoStop}%
	\bibitem [{\citenamefont {Andersen}(1975)}]{andersen75}%
	\BibitemOpen
	\bibfield  {author} {\bibinfo {author} {\bibfnamefont {O.~K.}\ \bibnamefont
			{Andersen}},\ }\href@noop {} {\bibfield  {journal} {\bibinfo  {journal}
			{Phys. Rev. B}\ }\textbf {\bibinfo {volume} {12}},\ \bibinfo {pages} {3060}
		(\bibinfo {year} {1975})}\BibitemShut {NoStop}%
	\bibitem [{\citenamefont {Takeda}\ and\ \citenamefont
		{K{\H{u}}bler}(1979)}]{takeda79}%
	\BibitemOpen
	\bibfield  {author} {\bibinfo {author} {\bibfnamefont {T.}~\bibnamefont
			{Takeda}}\ and\ \bibinfo {author} {\bibfnamefont {J.}~\bibnamefont
			{K{\H{u}}bler}},\ }\href@noop {} {\bibfield  {journal} {\bibinfo  {journal}
			{J. Phys. F}\ }\textbf {\bibinfo {volume} {9}},\ \bibinfo {pages} {661}
		(\bibinfo {year} {1979})}\BibitemShut {NoStop}%
	\bibitem [{\citenamefont {Hohenberg}\ and\ \citenamefont
		{Kohn}(1964)}]{hohenberg64}%
	\BibitemOpen
	\bibfield  {author} {\bibinfo {author} {\bibfnamefont {P.}~\bibnamefont
			{Hohenberg}}\ and\ \bibinfo {author} {\bibfnamefont {W.}~\bibnamefont
			{Kohn}},\ }\href@noop {} {\bibfield  {journal} {\bibinfo  {journal} {Phys.
				Rev.}\ }\textbf {\bibinfo {volume} {136}},\ \bibinfo {pages} {864} (\bibinfo
		{year} {1964})}\BibitemShut {NoStop}%
	\bibitem [{\citenamefont {Kohn}\ and\ \citenamefont {Sham}(1965)}]{kohn65}%
	\BibitemOpen
	\bibfield  {author} {\bibinfo {author} {\bibfnamefont {W.}~\bibnamefont
			{Kohn}}\ and\ \bibinfo {author} {\bibfnamefont {L.~J.}\ \bibnamefont
			{Sham}},\ }\href@noop {} {\bibfield  {journal} {\bibinfo  {journal} {Phys.
				Rev.}\ }\textbf {\bibinfo {volume} {140}},\ \bibinfo {pages} {1133} (\bibinfo
		{year} {1965})}\BibitemShut {NoStop}%
	\bibitem [{\citenamefont {Vosko}\ \emph {et~al.}(1980)\citenamefont {Vosko},
		\citenamefont {Wilk},\ and\ \citenamefont {Nusair}}]{vosko80}%
	\BibitemOpen
	\bibfield  {author} {\bibinfo {author} {\bibfnamefont {S.~H.}\ \bibnamefont
			{Vosko}}, \bibinfo {author} {\bibfnamefont {L.}~\bibnamefont {Wilk}},\ and\
		\bibinfo {author} {\bibfnamefont {M.}~\bibnamefont {Nusair}},\ }\href@noop {}
	{\bibfield  {journal} {\bibinfo  {journal} {Can. J. Phys.}\ }\textbf
		{\bibinfo {volume} {58}},\ \bibinfo {pages} {1200} (\bibinfo {year}
		{1980})}\BibitemShut {NoStop}%
	\bibitem [{\citenamefont {Painter}(1981)}]{painter81}%
	\BibitemOpen
	\bibfield  {author} {\bibinfo {author} {\bibfnamefont {G.~S.}\ \bibnamefont
			{Painter}},\ }\href@noop {} {\bibfield  {journal} {\bibinfo  {journal} {Phys.
				Rev. B}\ }\textbf {\bibinfo {volume} {24}},\ \bibinfo {pages} {4264}
		(\bibinfo {year} {1981})}\BibitemShut {NoStop}%
	\bibitem [{\citenamefont {Anisimov}\ \emph {et~al.}(1991)\citenamefont
		{Anisimov}, \citenamefont {Zaanen},\ and\ \citenamefont
		{Andersen}}]{anisimov91}%
	\BibitemOpen
	\bibfield  {author} {\bibinfo {author} {\bibfnamefont {V.~I.}\ \bibnamefont
			{Anisimov}}, \bibinfo {author} {\bibfnamefont {J.}~\bibnamefont {Zaanen}},\
		and\ \bibinfo {author} {\bibfnamefont {O.~K.}\ \bibnamefont {Andersen}},\
	}\href@noop {} {\bibfield  {journal} {\bibinfo  {journal} {Phys. Rev. B}\
		}\textbf {\bibinfo {volume} {44}},\ \bibinfo {pages} {943} (\bibinfo {year}
		{1991})}\BibitemShut {NoStop}%
	\bibitem [{\citenamefont {Solovyev}\ \emph {et~al.}(1996)\citenamefont
		{Solovyev}, \citenamefont {Hamada},\ and\ \citenamefont
		{Terakura}}]{solovyev96a}%
	\BibitemOpen
	\bibfield  {author} {\bibinfo {author} {\bibfnamefont {I.}~\bibnamefont
			{Solovyev}}, \bibinfo {author} {\bibfnamefont {N.}~\bibnamefont {Hamada}},\
		and\ \bibinfo {author} {\bibfnamefont {K.}~\bibnamefont {Terakura}},\
	}\href@noop {} {\bibfield  {journal} {\bibinfo  {journal} {Phys. Rev. B}\
		}\textbf {\bibinfo {volume} {53}},\ \bibinfo {pages} {7158} (\bibinfo {year}
		{1996})}\BibitemShut {NoStop}%
	\bibitem [{\citenamefont {Anisimov}\ \emph {et~al.}(1997)\citenamefont
		{Anisimov}, \citenamefont {Aryasetiawan},\ and\ \citenamefont
		{Linchtenstein}}]{anisimov97}%
	\BibitemOpen
	\bibfield  {author} {\bibinfo {author} {\bibfnamefont {V.~I.}\ \bibnamefont
			{Anisimov}}, \bibinfo {author} {\bibfnamefont {F.}~\bibnamefont
			{Aryasetiawan}},\ and\ \bibinfo {author} {\bibfnamefont {A.~I.}\ \bibnamefont
			{Linchtenstein}},\ }\href@noop {} {\bibfield  {journal} {\bibinfo  {journal}
			{J. Phys.: Condens. Matter}\ }\textbf {\bibinfo {volume} {9}},\ \bibinfo
		{pages} {767} (\bibinfo {year} {1997})}\BibitemShut {NoStop}%
	\bibitem [{\citenamefont {Takegami}\ \emph {et~al.}(2019)\citenamefont
		{Takegami}, \citenamefont {Nicola{\"i}}, \citenamefont {Koethe},
		\citenamefont {kasinathan}, \citenamefont {Kuo}, \citenamefont {Liao},
		\citenamefont {Tsuei}, \citenamefont {Panaccione}, \citenamefont {Offi},
		\citenamefont {Monaco}, \citenamefont {Brookes}, \citenamefont {Min{\'a}r},\
		and\ \citenamefont {Tjeng}}]{Takegami19}%
	\BibitemOpen
	\bibfield  {author} {\bibinfo {author} {\bibfnamefont {D.}~\bibnamefont
			{Takegami}}, \bibinfo {author} {\bibfnamefont {L.}~\bibnamefont
			{Nicola{\"i}}}, \bibinfo {author} {\bibfnamefont {T.~C.}\ \bibnamefont
			{Koethe}}, \bibinfo {author} {\bibfnamefont {D.}~\bibnamefont {kasinathan}},
		\bibinfo {author} {\bibfnamefont {C.~Y.}\ \bibnamefont {Kuo}}, \bibinfo
		{author} {\bibfnamefont {Y.~F.}\ \bibnamefont {Liao}}, \bibinfo {author}
		{\bibfnamefont {K.~D.}\ \bibnamefont {Tsuei}}, \bibinfo {author}
		{\bibfnamefont {G.}~\bibnamefont {Panaccione}}, \bibinfo {author}
		{\bibfnamefont {F.}~\bibnamefont {Offi}}, \bibinfo {author} {\bibfnamefont
			{G.}~\bibnamefont {Monaco}}, \bibinfo {author} {\bibfnamefont {N.~B.}\
			\bibnamefont {Brookes}}, \bibinfo {author} {\bibfnamefont {J.}~\bibnamefont
			{Min{\'a}r}},\ and\ \bibinfo {author} {\bibfnamefont {L.~H.}\ \bibnamefont
			{Tjeng}},\ }\href@noop {} {\bibfield  {journal} {\bibinfo  {journal} {Phys.\
				Rev. B}\ }\textbf {\bibinfo {volume} {99}},\ \bibinfo {pages} {165101}
		(\bibinfo {year} {2019})}\BibitemShut {NoStop}%
	\bibitem [{Note5()}]{Note5}%
	\BibitemOpen
	\bibinfo {note} {We have also obtained the same result on LaCoO$_3$ and other
		lanthanum oxides such as La$_{1-x}$Sr$_x$MnO$_3$ using $\sim$8 keV photon
		energy at BL09XU and 47XU of SPring-8 while Ref.~\onlinecite{Takegami19}
		employed 6.5 keV photon energy.}\BibitemShut {Stop}%
\end{thebibliography}
%

%

\end{document}